\documentclass[12pt]{article}

\global\arraycolsep=1pt
\usepackage{amsmath,amssymb}

\numberwithin{equation}{section}
\newcommand{\beq}{\begin{equation}}
\newcommand{\eeq}{\end{equation}}
\newcommand{\beqa}{\begin{eqnarray}}
\newcommand{\eeqa}{\end{eqnarray}}

\renewcommand{\thefootnote}{\fnsymbol{footnote}}

\begin{document}

\begin{flushright}
December, 2008 \\
OCU-PHYS 309 \\
\end{flushright}
\vspace{5mm}

\begin{center}
{\bf\Large
Separability of Gravitational Perturbation \\
in Generalized Kerr-NUT-de Sitter Spacetime
}
\end{center}

\begin{center}

\vspace{10mm}

Takeshi Oota$^a$\footnote{
\texttt{toota@sci.osaka-cu.ac.jp}
} and
Yukinori Yasui$^b$\footnote{
\texttt{yasui@sci.osaka-cu.ac.jp}
}

\vspace{10mm}

\textit{
${{}^a}$Osaka City University
Advanced Mathematical Institute (OCAMI)\\
3-3-138 Sugimoto, Sumiyoshi,
Osaka 558-8585, JAPAN
}

\vspace{5mm}

\textit{
${}^b$Department of Mathematics and Physics, Graduate School of Science,\\
Osaka City University\\
3-3-138 Sugimoto, Sumiyoshi,
Osaka 558-8585, JAPAN
}

\vspace{5mm}

\end{center}
\vspace{8mm}

\begin{abstract}
Generalized Kerr-NUT-de Sitter spacetime is 
the most general spacetime
which admits a rank-2 closed conformal Killing-Yano tensor.
It contains the
higher-dimensional Kerr-de Sitter black holes 
with partially equal angular momenta.
We study the separability of gravitational perturbations in 
the generalized Kerr-NUT-de Sitter spacetime.
We show that a certain type of tensor perturbations 
admits the separation of variables.
The linearized perturbation equations for the Einstein condition
are transformed into the ordinary differential equations
of Fuchs type.
\end{abstract}

\vspace{25mm}

\newpage

\renewcommand{\thefootnote}{\arabic{footnote}}
\setcounter{footnote}{0}


\section{Introduction}

The higher-dimensional Kerr-NUT-de Sitter metric was 
constructed by Chen-L\"u-Pope \cite{CLP}.
The metric is the most general known solution 
describing the rotating asymptotically de Sitter black hole spacetime 
with NUT parameters. It has been shown in \cite{FK,KF} 
that the Kerr-NUT-de Sitter spacetime 
has a rank-2 closed conformal Killing-Yano~(CKY) tensor \cite{tac}.
This tensor generates the tower of Killing-Yano and Killing tensors, 
which implies complete integrability of geodesic
equations \cite{PKVK,KKPV}
and complete separation of variables for the Hamilton-Jacobi \cite{FK}, 
Klein-Gordon \cite{FK} and Dirac equations \cite{OY2}
(for reviews, see \cite{flor08,FK08,kub}).  
Furthermore, it was proved that the Kerr-NUT de Sitter
spacetime is the only spacetime when the eigenvalues of
the CKY tensor are functionally independent \cite{HOY,HOY2,KFK08}.

Recently, we have obtained the most general metric 
admitting a rank-2 closed CKY tensor \cite{HOY3,HOY4}.
The CKY tensor generally has the constant eigenvalues 
besides the functionally independent eigenvalues.
Associated with these constant eigenvalues the spacetime admits
K\"ahler manifolds of the same dimension as the multiplicity of them.
Then, the metric may be locally written as 
a Kaluza-Klein metric on the bundle over the K\"ahler
manifolds whose fibers are Kerr-NUT-de Sitter spacetimes.
We call such a spacetime the generalized Kerr-NUT-de Sitter spacetime.

Important examples are given by a special class 
of Kerr-de Sitter metrics found by
Gibbons-L\"u-Page-Pope \cite{GLPP1,GLPP2}.
In $D$ dimensions the Kerr-de Sitter metric 
has $[(D-1)/2]$ angular momenta. When some of them are equal,
the CKY tensor has  constant eigenvalues. 
Then, the coordinates used in \cite{CLP}, 
which are based on the eigenvalues of the CKY tensor,  
are not effective to express the metric
since the constant eigenvalues do not work as coordinates. 
Actually, such a spacetime belongs not to the Kerr-NUT-de Sitter
but to the generalized
Kerr-NUT-de Sitter spacetime. 
In particular the odd-dimensional Kerr-de Sitter metric for which all
angular momenta are equal describes a cohomogeneity one spacetime. 

In this paper we study the separability of gravitational 
perturbations in the generalized Kerr-NUT-de Sitter
spacetime. In four dimensions there exists a master equation 
describing the perturbations of the rotating black holes
\cite{teu1,teu2}.
However, in higher-dimensional spacetimes no one 
has succeeded in finding such an equation. Recently, there was 
a certain progress in this problem 
\cite{KI03a,IK03,KI03b,KLR,KZ0703,kod0711,kod0712,KMIS,IKKMSZ,
MS,KMSZ,KKZ,mur}.
In \cite{KLR} the perturbation of 
odd-dimensional Myers-Perry black hole 
with equal angular momenta was studied. 
They  found a class of perturbations for which
the equations of motion reduce to a single radial equation. 
We extend their analysis to the case of
the generalized Kerr-NUT-de Sitter spacetime.  
Our main result shows that the tensor type gravitational 
perturbations for the generalized Kerr-NUT-de Sitter
spacetime admit the separation of variables.

This paper is organized as follows.
In section 2 we briefly describe the properties 
of the generalized Kerr-NUT-de Sitter metric.
In section 3 the solutions of the Einstein condition
are briefly reviewed. To illustrate the solutions,
we give a subfamily of the solutions which represents
the higher-dimensional Kerr-de Sitter black holes
with partially equal angular momenta (and with some zero
angular momenta). 
This provides important examples of our formulation. 
In section 4 we show the separability of the tensor 
type perturbations. It should be emphasised that
the separability is a rather non-trivial consequence 
of the detailed structure of the spacetime.
A recent study of the separability in the 
gravitational perturbations was restricted to the case of
all angular momenta equal, where the spacetime 
is of cohomogeneity one. In our case, the equations can be
separated in the situation where the black holes 
have angular momenta of plural different values.
In section 5 we summarize the results and comment 
on open questions. 
In Appendices A and B, we present the connection 1-forms 
and the curvature 2-forms explicitly.
These quantities are essential to our calculations.
Some of the results presented here are already 
available in \cite{HOY3,HOY4}. 
However, we attempt to make
this paper as self-contained as possible.
We also provide Appendices C and D,
which contain some details of our calculations.  
We obtain an explicit coordinate 
transformation from the higher-dimensional
Kerr-de Sitter black hole with partially 
equal angular momenta to the generalized Kerr-NUT-de Sitter
spacetime. In Appendix E, we present non-zero components of the Lichnerowicz operator.

\section{Generalized Kerr-NUT-de Sitter spacetime}

In this section, we review the metric on the
generalized Kerr-NUT-de Sitter spacetime which
admits a rank-2 closed CKY tensor. 

We start with the $2n$-dimensional Kerr-NUT-de Sitter metric
found by Chen-L\"{u}-Pope \cite{CLP}. The metric takes the form
\begin{equation} \label{KNdSmet}
g^{(2n)} = \sum_{\mu=1}^{n} \frac{d x_{\mu}^2}{Q_\mu (x)}+
\sum_{\mu=1}^{n} Q_{\mu}(x) \left( 
\sum_{k=0}^{n-1} \sigma_{k}(\hat{x}_{\mu}) 
d \psi_k \right)^2,
\end{equation}
where  $Q_{\mu}$ is defined by
\begin{equation}
Q_{\mu} = \frac{Y_{\mu}}{U_{\mu}}, \qquad
U_{\mu}=\prod_{\stackrel{\scriptstyle \nu=1}
{(\nu \ne \mu)}}^{n}
(x_{\mu}^2-x_{\nu}^2)
\end{equation}
with a function $Y_{\mu} = Y_{\mu}(x_{\mu})$ depending
only on $x_{\mu}$. The coordinates $\psi_k$ give
the Killing vectors $\partial/\partial \psi_k$ 
$(k=0,1,\dotsc, n-1)$.
The symbol $\sigma_k(\hat{x}_{\mu})$
are the $k$-th elementary symmetric functions of 
$\{x_{\nu}^2 ; \nu \neq \mu\}$:
\begin{equation}
\prod_{\stackrel{\scriptstyle \nu=1}{(\nu \ne \mu)}}^{n}
(t-x_{\nu}^2)=\sigma_0(\hat{x}_{\mu}) t^{n-1}
-\sigma_1(\hat{x}_{\mu}) t^{n-2}
+ \cdots + (-1)^{n-1} \sigma_{n-1}(\hat{x}_{\mu}).
\end{equation}

The spacetime admits a rank-$2$ non-degenerate closed CKY
tensor \cite{FK,KF}. Without assuming the non-degeneracy
the classification of higher-dimensional spacetimes $(M,g)$ 
with a rank-2 closed CKY tensor  was obtained
in \cite{HOY3,HOY4}. 
We call such spacetimes the generalized Kerr-NUT-de Sitter
spacetimes. They have a bundle structure;
the fiber space is the $2n$-dimensional Kerr-NUT-de Sitter spacetime
and the base space is a product space
\begin{equation}
B = M^{(1)} \times M^{(2)} \times \dotsm
\times M^{(N)} \times M^{(0)},
\end{equation}
where the manifolds $M^{(i)}$ $(i=1,2,\dotsc, N)$ are $2m_i$-dimensional
K\"{a}hler manifolds with metrics $g^{(i)}$,
and $M^{(0)}$ is an $m^{(0)}$-dimensional Riemann manifold
with a metric $g^{(0)}$.
The dimension $D$ of the generalized 
Kerr-NUT-de Sitter spacetime is given by
\beq
D = 2 n + 2 | m | + m^{(0)}, \qquad
|m|:= \sum_{i=1}^N m_i.
\eeq
The coordinates $(x_{\mu}, \psi_k)$ ($\mu=1,2,\dotsc, n$ and
$k=0,1,\dotsc, n-1$) are coordinates on the 
$2n$-dimensional Kerr-NUT-de Sitter fiber space.
The fiber metric is twisted by the K\"{a}hler forms $\omega^{(i)}$
corresponding to the metric $g^{(i)}$,
i.e., the $1$-form $d \psi_k$ in \eqref{KNdSmet}
is replaced by a $1$-form $\theta_k$.
Locally, $\omega^{(i)} = dA^{(i)}$ and so we can write it as
\begin{equation} \label{eqth}
\theta_k = d \psi_k - 2 \sum_{i=1}^N (-1)^{(n-k)} \xi_i^{2(n-k)-1} A^{(i)},
\qquad k=0,1,\dotsc, n-1.
\end{equation}

The metric on the generalized Kerr-NUT-de Sitter
spacetime
is of the form
\begin{equation} \label{gCKY}
g = 
\sum_{\mu=1}^{n} \frac{d x_{\mu}^2}{P_\mu (x)}+
\sum_{\mu=1}^{n} P_{\mu}(x) \left( 
\sum_{k=0}^{n-1} \sigma_{k}(\hat{x}_{\mu}) 
\theta_k \right)^2
+ \sum_{i=1}^{N}\prod_{\mu=1}^{n}
(x_{\mu}^2-\xi_i^2)g^{(i)}
+\sigma_n g^{(0)},
\end{equation}
and the CKY tensor is written as
\begin{equation} \label{cCKY}
h=\sum_{\mu=1}^n x_{\mu} dx_{\mu} \wedge 
\left(\sum_{k=0}^{n-1} \sigma_{k}(\hat{x}_{\mu})
\theta_k \right)
+\sum_{i=1}^N \xi_i \prod_{\mu=1}^{n}
(x_{\mu}^2-\xi_i^2)\omega^{(i)},
\end{equation}
where the function $P_{\mu}$ is defined by
\begin{equation} \label{Pmu}
P_{\mu}(x)=\frac{X_{\mu}}{\displaystyle
(x_{\mu})^{m^{(0)}} 
\prod_{i=1}^{N}(x_{\mu}^2-\xi_i^2)^{m_i}U_{\mu}},
\qquad
U_{\mu}=\prod_{\stackrel{\scriptstyle \nu=1}
{(\nu \ne \mu)}}^{n}
(x_{\mu}^2-x_{\nu}^2)
\end{equation}
with an arbitrary function $X_{\mu}=X_{\mu}(x_{\mu})$ 
depending only on $x_{\mu}$.
In these expressions 
\begin{enumerate}
\item[(a)] the coordinates $x_{\mu}$ 
($\mu=1,\cdots, n$) and the parameters $\xi_i$ 
($i=1,\cdots, N$) are the non-constant eigenvalues 
and the non-zero constant eigenvalues of $h$, respectively.
\item[(b)] the dimension $2m_i$ of the K\"{a}hler 
manifold $M^{(i)}$
is equal to multiplicity of non-zero constant eigenvalue
$\xi_i$, and the dimension $m^{(0)}$ of the Riemann manifold
$M^{(0)}$ is equal to multiplicity of zero eigenvalue.
\item[(c)] 
for $m^{(0)}=1$ the last term in \eqref{gCKY} can take the
special form:
\begin{equation}\label{g0sp}
\sigma_n g^{(0)} ~~\longrightarrow~~ \sigma_n g^{(0)}_{\mathrm{special}} 
=\frac{c}{\sigma_n} \left( \sum_{k=0}^{n} \sigma_k
\theta_k \right)^2
\end{equation}
with a constant $c$. Here $\sigma_k$ is the $k$-th elementary
symmetric functions of $\{ x_{1}^2, \dotsc, x_{n}^2 \}$:
\beq
\prod_{\nu=1}^{n}(t-x_{\nu}^2)
=\sigma_0t^n-\sigma_1 t^{n-1}+ \cdots + (-1)^n \sigma_n.
\eeq
We note that the odd-dimensional Kerr-NUT-de Sitter spacetime
belongs to the special type with $N=0$.
\item[(d)] the 1-forms $\theta_k$ satisfy
\begin{equation}
d\theta_k+ 2\sum_{i=1}^N (-1)^{(n-k)} 
\xi_i^{2n-2k-1} \omega^{(i)}=0, \qquad
k=0,1,\dotsc, n-1+\varepsilon,
\end{equation}
where $\omega^{(i)}$ is a K\"ahler form on $M^{(i)}$ and
$\varepsilon=0$ for the general type and $\varepsilon=1$
for the special type.
\end{enumerate}

\section{Einstein condition and black hole solutions}

For the generalized Kerr-NUT-de Sitter spacetime $(M,g)$ 
the Einstein condition
\begin{equation}
Ric(g)=\Lambda g
\end{equation}
was explicitly solved in \cite{HOY3}. 
In appendices A and B we give explicit forms
of the spin connection $1$-forms and the Riemann curvature
$2$-forms which were omitted in \cite{HOY3}.

The result is as follows: 
the base metrics $g^{(0)}$ and 
$g^{(i)}$ are Einstein and
the function $X_{\mu}$ in \eqref{Pmu} takes the form
\begin{equation} \label{solX}
X_{\mu}(x_{\mu})=x_{\mu} \left( 
d_{\mu}+\int \chi(x_{\mu}) \, x_{\mu}{}^{m^{(0)}-2} 
\prod_{i=1}^{N}(x_{\mu}^2-\xi_i^2)^{m_i} dx_{\mu} \right),
\end{equation}
where
\begin{equation} \label{chi}
\chi(x)=\sum_{i=-\varepsilon}^n 
\alpha_i x^{2i},~~~\alpha_n=-\Lambda,
\end{equation}
\begin{flushleft}
(a) general type $(\varepsilon=0)$
\end{flushleft}
\begin{equation} \label{al0}
\alpha_0=(-1)^{n-1} \lambda^{(0)},
\end{equation}
\begin{flushleft}
(b) special type ($m^{(0)}=1$ and $\varepsilon=1$)
\end{flushleft}
\begin{equation}
\alpha_0=(-1)^{n-1}2c \sum_{j=1}^{N} 
\frac{m_j}{\xi_j^2},~~~\alpha_{-1}=(-1)^{n-1}2 c .
\end{equation}
Here, $\alpha_i$ and $d_{\mu}$ are constants. 
The $\lambda^{(0)}$ in \eqref{al0} is 
a cosmological constant of $g^{(0)}$, 
and cosmological constants $\lambda^{(i)}$ of $g^{(i)}$ 
are given by
\begin{equation} \label{cosmi}
\lambda^{(i)}=(-1)^{n-1} \chi(\xi_i).
\end{equation}

To illustrate the above solutions \eqref{solX}, 
in the following subsections 
\ref{secsp} and \ref{secgen}, 
we consider a subfamily of the solutions
which represents the Kerr-de Sitter black holes with
partially equal angular momenta (and with some zero
angular momenta).

\subsection{Special type $(D=2n+2|m|+1)$}
\label{secsp}

First we consider a particular subset of the special case 
$(m^{(0)}=1, \varepsilon=1)$.
Let us set the number of base K\"{a}hler manifolds 
$N=n$. We choose $n$ K\"{a}hler manifolds $M^{(i)}$ to be 
the complex projective spaces $\mathbb{CP}^{m_i}$.
The base space $B$ is
\beq
B = M^{(1)} \times M^{(2)} \times \dotsm \times M^{(n)}
= \mathbb{CP}^{m_1} \times \mathbb{CP}^{m_2}
\times \dotsm \times \mathbb{CP}^{m_n},
\eeq
and the fiber will be chosen as a $(2n+1)$-dimensional
Kerr-de Sitter spacetime.

We restrict various parameters as follows
\begin{equation}
D= 2n + 2|m|+1, \qquad |m|= \sum_{i=1}^n m_i,
\qquad
\Lambda = (D-1)\lambda = 2(n+|m|) \lambda.
\end{equation}
The metric of special type with $N=n$ is given by
\begin{equation} \label{spsp}
g = \sum_{\mu=1}^n \frac{dx_{\mu}^2}{P_{\mu}} 
+ \sum_{\mu=1}^n P_{\mu}
\left[ \sum_{k=0}^{n-1} 
\sigma_k( \hat{x}_{\mu}) \theta_k \right]^2 
+ \frac{c}{\sigma_n}
\left[ \sum_{k=0}^{n} \sigma_k \theta_k \right]^2
+ \sum_{i=1}^n \prod_{\mu=1}^n (x_{\mu}^2 - \xi_i^2) g^{(i)}.
\end{equation}
We choose the functions $P_{\mu}$ and the constant $c$
as follows:
\begin{equation}
P_{\mu} = \frac{X_{\mu}(x_{\mu})}
{\displaystyle x_{\mu} \prod_{i=1}^n
(x_{\mu}^2 - \xi_i^2)^{m_i} U_{\mu}},
\qquad
c = - \prod_{i=1}^n \xi_i^2,
\end{equation}
\begin{equation} \label{Xsolsp}
X_{\mu}(x_{\mu}) = x_{\mu}\left( \tilde{d}_{\mu}
- \left(1 + \lambda x_{\mu}^2 \right) x_{\mu}^{-2}
\prod_{i=1}^n ( x_{\mu}^2 - \xi_i^2)^{m_i+1} \right).
\end{equation}
Here $g^{(i)}$ is the Fubini-Study metric on 
$\mathbb{CP}^{m_i}$ with the cosmological constant $\lambda^{(i)}$.
The constants $\tilde{d}_{\mu}$ correspond 
to the mass $M$ and the NUT parameters. 
We set all NUT parameters to be zero:
\beq
\tilde{d}_{\mu} = (-1)^{(1/2)(D-1)-1} 2M \delta_{\mu,n}.
\eeq
We can show that the above metric \eqref{spsp} represents 
the $D=2n'+1$ dimensional Kerr-de Sitter
metric \cite{GLPP1,GLPP2} with partially equal
angular momenta:
for $n'=n+|m|$, among $n'$ angular momenta $\{a_I\}$, 
$m_i+1$ $a$'s are taken equal to $\xi_i$ $(i=1,2,\dotsc, n)$.
Here the non-zero constants $\xi_i$ are assumed to be all different:
$\xi_i \neq \xi_j$ $(\forall i \neq \forall j)$.
See Appendix C for details.

The function $X_{\mu}(x_{\mu})$ \eqref{Xsolsp}
is indeed the special case of \eqref{solX}.
The corresponding function $\chi(x)$ \eqref{chi} in this case
is given by
\begin{equation}
\begin{split} \label{chisp}
\chi(x) = \sum_{i=-1}^{n} \alpha_i x^{2i} 
&= -2 \sum_{i=1}^n (m_i+1) ( 1 + \lambda \xi_i^2) 
\prod_{\stackrel{\scriptstyle j=1}{(j \neq i)}}^n 
(x^2 - \xi_j^2) \cr
&\ \ \ + 2 \left( \frac{1}{x^2} 
- ( n + |m|) \lambda \right)
\prod_{i=1}^n (x^2 - \xi_i^2).
\end{split}
\end{equation}
Explicit form of some parameters $\alpha_i$ can be read off:
\begin{equation}
\alpha_{-1} = (-1)^n 2 \prod_{i=1}^n \xi_i^2, \qquad
\alpha_0 = (-1)^n 2 \prod_{i=1}^n \xi_i^2
\left( \sum_{j=1}^n \frac{m_j}{\xi_j^2} \right), \qquad
\alpha_n = -2 ( n + |m|) \lambda.
\end{equation}
The cosmological constant $\lambda^{(i)}$ of $g^{(i)}$ is given by
\begin{equation}
\lambda^{(i)} = (-1)^{n-1} \chi(\xi_i) 
= (-1)^n 2 (m_i+1) (1 + \lambda \xi_i^2)
\prod_{\stackrel{\scriptstyle j=1}{(j \neq i)}}^n
( \xi_i^2 - \xi_j^2).
\end{equation}

\subsection{General type $(D=2n+2|m|+m^{(0)})$}
\label{secgen}

Next we consider a particular subset 
of the general type $(\varepsilon=0)$.
Let the number of base K\"{a}hler manifolds be
$N=n-1$, and take 
$M^{(i)} = \mathbb{CP}^{m_i}$ $(i=1,2,\dotsc, n-1)$.
Also we take $M^{(0)}$ to be an $m^{(0)}$-dimensional sphere
$S^{m^{(0)}}$. The base space $B$ is 
\beq
B = M^{(1)} \times M^{(2)} \times \dotsm \times M^{(n-1)}
\times M^{(0)}
= \mathbb{CP}^{m_1} \times \mathbb{CP}^{m_2}
\times \dotsm \times \mathbb{CP}^{m_{n-1}}
\times S^{m^{(0)}}.
\eeq
We will choose the fiber as a $2n$-dimensional Kerr-de Sitter spacetime.
Hence
\beq
D = 2n + 2|m| + m^{(0)}, 
\qquad 
|m| = \sum_{i=1}^{n-1} m_i, \qquad
\Lambda = (D-1)\lambda.
\end{equation}
The metric of general type with $N=n-1$ is given by
\begin{equation} \label{evng}
g = \sum_{\mu=1}^n \frac{dx_{\mu}^2}{P_{\mu}}
+ \sum_{\mu=1}^n P_{\mu}
\left[ \sum_{k=0}^{n-1} \sigma_k(\hat{x}_{\mu}) \theta_k \right]^2
+ \sum_{i=1}^{n-1} \prod_{\mu=1}^n (x_{\mu}^2 - \xi_i^2) g^{(i)}
+ \sigma_n g^{(0)}.
\end{equation}
Here $g^{(i)}$ is the Fubini-Study metric on $\mathbb{CP}^{m_i}$
with the cosmological constant $\lambda^{(i)}$
and $g^{(0)}$ is the standard metric on the sphere $S^{m^{(0)}}$
with the cosmological constant $\lambda^{(0)}$.
The functions $P_{\mu}$ are chosen as follows:
\beq
P_{\mu} = \frac{X_{\mu}(x_{\mu})}{\displaystyle (x_{\mu})^{m^{(0)}}
\prod_{i=1}^{n-1} ( x_{\mu}^2 - \xi_i^2)^{m_i} U_{\mu}}, \qquad
\mu=1,2,\dotsc, n,
\eeq
where
\beq
X_{\mu}(x_{\mu}) 
= x_{\mu} \left( \tilde{d}_{\mu}
- (1+\lambda x_{\mu}^2) x_{\mu}^{m^{(0)}-1}
\prod_{i=1}^{n-1} ( x_{\mu}^2 - \xi_i^2)^{m_i+1} \right),
\eeq
with zero NUT parameters
\beq
\tilde{d}_{\mu}
= \begin{cases}
(-1)^{(1/2)D-1} 2M i \delta_{\mu,n} & (D \ \mbox{even}) \cr
(-1)^{(1/2)(D-1)-1} 2M \delta_{\mu,n} & (D \ \mbox{odd}).
\end{cases}
\eeq
This metric \eqref{evng} represents
the general Kerr-de Sitter metric \cite{GLPP1,GLPP2}
with partially equal angular momenta and 
with some zero angular momenta.
See Appendix C.1 and D for details. 

In this case, the function \eqref{chi} is given by
\beq
\begin{split}
\chi(x) = \sum_{i=0}^{n} \alpha_i x^{2i} 
&= -2 (1 + \lambda x^2) \sum_{i=1}^{n-1} (m_i + 1) \xi_i^2
\prod_{\stackrel{\scriptstyle j=1}{(j \neq i)}}^{n-1}
(x^2 - \xi_j^2) \cr
& \ \ \ + \Bigl( 2 - (D-1) (1 + \lambda x^2) \Bigr)
\prod_{j=1}^{n-1} ( x^2 - \xi_j^2).
\end{split}
\eeq
Here
\beq
\alpha_0 = (-1)^{n-1} \lambda^{(0)}, \qquad
\lambda^{(0)} = -(m^{(0)} -1) \prod_{i=1}^{n-1} \xi_i^2,
\eeq
\beq
\lambda^{(i)} = (-1)^{n-1} \chi(\xi_i)
= (-1)^n 2  (m_i + 1) ( 1 + \lambda \xi_i^2) \xi_i^2
\prod_{\stackrel{\scriptstyle j=1}{(j \neq i)}}^{n-1}
(\xi_i^2 - \xi_j^2).
\eeq

Note that the standard metric $d\Omega_{(m^{(0)})}^2$
on $S^{m^{(0)}}$ with 
unit radius has the cosmological constant $(m^{(0)}-1)$.
Hence,
\beq
g^{(0)} = - \left( \prod_{i=1}^{n-1} \xi_i^{-2} \right) 
d \Omega_{(m^{(0)})}^2.
\eeq

\section{Separability of gravitational perturbation}

In this section we study a linear perturbation 
$g_{AB} \rightarrow g_{AB}+h_{AB}$ 
of the generalized Kerr-NUT-de Sitter metric $g=(g_{AB})$.
We assume that the background metric satisfies the Einstein condition. 
Under the traceless and transverse conditions
\begin{equation}
g^{AB} h_{AB}=0,~~~\nabla^A h_{AB}=0,
\end{equation}
the linearized Einstein equation is given by
\begin{equation} \label{linEin}
\Delta_L h_{AB}=2 \Lambda h_{AB},
\end{equation} 
where the Lichnerowicz operator $\Delta_L$ is defined by
\begin{equation}
\Delta_{L} h_{AB}=-\nabla^C \nabla_C h_{AB}-2 R_{ACBD} h^{CD}
+2 \Lambda h_{AB}.
\end{equation}
Using the orthonormal frame $\{ e_A \}$, we have
\begin{equation}
\nabla_A h_{BC}=e_A(h_{BC})-h_{DC} \omega_{DB}(e_A)
-h_{BD} \omega_{DC}(e_A),
\end{equation}
\begin{eqnarray}
\nabla_{C} \nabla_{C} h_{AB}&=&e_{C}(\nabla_C h_{AB})
-(\nabla_D h_{AB}) \omega_{DC}(e_C)\nonumber\\
&-&(\nabla_C h_{DB}) \omega_{DA}(e_C)
-(\nabla_C h_{AD}) \omega_{DB}(e_C).
\end{eqnarray}
Before describing details of perturbation,
we summarize the index labeling.
Recall that the spacetime has a bundle structure.
We use

(i) $\mu$ or $n+\mu$ ($\mu=1,2,\dotsc, n$)
for the Kerr-NUT-de Sitter fiber spacetime,

(ii) $(\hat{\alpha}, i) = ( \alpha,i)$ or $(m_i + \alpha, i)$
($\alpha=1,\dotsc, m_i$) for the $i$-th base K\"{a}hler manifold
$M^{(i)}$,

(iii-1) $a$ $(a=1,\dotsc, m^{(0)})$ for the general type
base Riemann manifold $M^{(0)}$,

(iii-2) $2n+1$ for the special type base Riemann manifold $M^{(0)}$.

The components $h_{AB}$ can be classified into scalar, 
vector and tensor components
according to the coordinate transformations of the base manifolds
$M^{(0)}$ and $M^{(i)}$.
Here we consider the perturbation of the base manifolds
$B = M^{(1)} \times M^{(2)} \times \dotsm \times M^{(N)}
\times M^{(0)}$. We call such perturbation a tensor perturbation
according to \cite{KLR}.
We will show later that the tensor components are decoupled
from other components (see below equation \eqref{DH2}).
Hence, one can consistently 
require that

(a) no scalar component
\begin{eqnarray} \label{noscal}
 h_{\mu \nu}&=&h_{\mu, n+\nu}=h_{n+\mu,n+\nu}=0,\\
h_{\mu,2n+1}&=&h_{n+\mu, 2n+1}=h_{2n+1,2n+1}=0, \nonumber
\end{eqnarray}

(b) no vector component
\begin{eqnarray} 
 h_{\mu,(\hat{\alpha},i)}&=&h_{n+\mu,(\hat{\alpha},i)}
=h_{2n+1,(\hat{\alpha},i)}=0,\\
 h_{\mu a}&=&h_{n+\mu,a}=0.\nonumber
 \end{eqnarray}
These conditions (a) and (b) mean that we do not
perturb the fiber metric and keep the bundle structure.

For simplicity, 
we further impose the 
following conditions on the tensor components:
\begin{equation}
\sum_{\hat{\alpha}=1}^{2 m_i} 
h_{(\hat{\alpha},i),(\hat{\alpha},i)}=
\sum_{\alpha=1}^{m_i}(h_{(\alpha,i),(\alpha,i)}
+h_{(m_i+\alpha,i),(m_i+\alpha,i)})=0~~~~\mbox{for each}~i,
\end{equation}
\begin{equation}
\sum_{a=1}^{m^{(0)}} h_{aa}=0
\end{equation}
and
\begin{equation}
h_{(\hat{\alpha},i),(\hat{\beta},j)}=0~~~
\mbox{for}~~i \ne j,~~~
h_{(\hat{\alpha},i),a}=0.
\end{equation}
Now, the traceless condition is automatically satisfied  
and the transverse condition reduces to
\begin{eqnarray}
\mathcal{D}^{(i)}_{\hat{\alpha}} 
h_{(\hat{\alpha},i),(\hat{\beta},i)}&:=&
\bar{e}^{(i)}_{\hat{\alpha}} (h_{(\hat{\alpha},i),(\hat{\beta},i)})
-h_{(\hat{\gamma},i),(\hat{\beta},i)}
\tilde{\omega}_{(\hat{\gamma},i),(\hat{\alpha},i)}
(\tilde{e}^{(i)}_{\hat{\alpha}})-
h_{(\hat{\alpha},i),(\hat{\gamma},i)}
\tilde{\omega}_{(\hat{\gamma},i),(\hat{\beta},i)}
(\tilde{e}^{(i)}_{\hat{\alpha}})\nonumber\\
&=& 0, \\
D^{(0)}_a h_{ab}&:=& \tilde{e}_a (h_{ab})-h_{cb} 
\tilde{\omega}_{ca}(\tilde{e}_a)
-h_{ac} \tilde{\omega}_{cb}(\tilde{e}_a) \nonumber\\
&=& 0,
\end{eqnarray}
where $\mathcal{D}^{(i)}$ is the gauge-covariant derivative 
on the K\"ahler-Einstein manifold $M^{(i)}$
and $D^{(0)}$ the covariant derivative on the 
Einstein manifold $M^{(0)}$. It should be noticed that
$\mathcal{D}^{(i)}$ includes the 1-form $A^{(i)}$ given 
by the K\"ahler form 
$\omega^{(i)}=d A^{(i)}$(see \eqref{gbei} and \eqref{sbei}).

\subsection{General type}

Now, we show that the equation \eqref{linEin} allows
a separation of variables for the tensor components
\begin{equation} \label{sephi}
h_{(\hat{\alpha},i),(\hat{\beta},i)}
=\left( \prod_{\mu=1}^{n} A^{(i)}_{\mu}(x_{\mu})
\prod_{k=0}^{n-1} e^{i N_k \psi_k} \right)
H^{(i)}_{\hat{\alpha} \hat{\beta}}(y_I^{(i)}) 
\prod_{\stackrel{\scriptstyle j=1}{(j \ne i)}}^{N} K^{(j)}(y_J^{(j)})
K^{(0)}(z_M),
\end{equation}
\begin{equation} \label{seph0}
h_{ab}=\left( \prod_{\mu=1}^{n} B_{\mu}(x_{\mu}) 
\prod_{k=0}^{n-1} e^{i N_k \psi_k} \right)
 \prod_{i=1}^{N} K^{(i)}(y_I^{(i)})H^{(0)}_{ab}(z_M),
\end{equation}
where $H^{(i)}_{\hat{\alpha} \hat{\beta}}(y_I^{(i)})$ and 
$H^{(0)}_{ab}(z_M)$ are tensor components
on $M^{(i)}$ and
$M^{(0)}$ respectively. $K^{(i)}(y_I^{(i)})$ and $K^{(0)}(z_M)$ 
are scalar functions on $M^{(i)}$ and $M^{(0)}$. Also,
$\{y_I^{(i)}; I=1,\cdots,2m_i \}$ and 
$\{z_M; M=1, \cdots, m^{(0)} \}$ represent 
the local coordinates on these spaces. 
Scalar components of \eqref{linEin} 
are trivially satisfied by the condition \eqref{noscal}.
We require the equation
\begin{equation} \label{DH1}
\mathcal{D}^{(i)}_{\beta} H^{(i)}_{m_i+\beta,\hat{\alpha}}-
\mathcal{D}^{(i)}_{m_i+\beta} H^{(i)}_{\beta,\hat{\alpha}}=0
\end{equation}
together with the transverse conditions:
\begin{equation} \label{DH2}
\mathcal{D}^{(i)}_{\hat{\alpha}} 
H^{(i)}_{\hat{\alpha} \hat{\beta}}
=0,~~~~D^{(0)}_a H^{(0)}_{ab}=0.
\end{equation}
This is a consequence of the vector component \eqref{gveccomp}. 
Thus the tensor components are decoupled from the scalar
and vector components.
It should be noticed that
from \eqref{gbei} the derivative $\bar{e}^{(i)}_{\hat{\alpha}}$ 
in the gauge-covariant derivative is given by
\begin{equation}
\bar{e}^{(i)}_{\hat{\alpha}}
:=\tilde{e}^{(i)}_{\hat{\alpha}}+i n_i A^{(i)}_{\hat{\alpha}},
\end{equation}
where
\begin{equation} \label{cni}
n_i=2 \sum_{k=0}^{n-1}(-1)^{n+k}\xi_i^{2(n-k)-1} N_k.
\end{equation} 
If we choose the Killing coordinates $\psi_i$ suitably,
then the charge $n_i$ is proportional to an integer, which
is interpreted as the first Chern number of the line bundle
over the K\"{a}hler-Einstein manifold $M^{(i)}$.

Let us evaluate the tensor components.
The non-zero tensor components of the Lichnerowicz operator
$\Delta_L h_{AB}$ are given in Appendix E.1.
The $\Box^{(F)}$-part was already studied 
in the separability of the Klein-Gordon equation 
on the Kerr-NUT-de Sitter background \cite{FKK}. 
Using the vector fields \eqref{gdualv} and the identity
\begin{equation}
\frac{1}{\prod_{\mu=1}^n (x_{\mu}^2-\xi_i^2)}
=(-1)^{n+1}\sum_{\mu=1}^n 
\frac{1}{U_{\mu}(x_{\mu}^2-\xi_i^2)},
\end{equation}
we find that the equation \eqref{linEin} takes the form
\begin{equation} \label{UG}
\sum_{\mu=1}^{n} \frac{1}{U_{\mu}} 
G^{(i)}_{\hat{\alpha} \hat{\beta}}(x_{\mu};y_I^{(i)})=0,~~~
\sum_{\mu=1}^{n} \frac{1}{U_{\mu}} G_{ab}(x_{\mu};z_M)=0,~~~
\end{equation}
where $G^{(i)}_{\hat{\alpha} \hat{\beta}}$ and $G_{ab}$
include the single coordinate
$x_{\mu}$. The explicit forms are given by
\begin{eqnarray} \label{exG}
G^{(i)}_{\alpha \beta}
&=&L^{(i)}(x_{\mu}) H^{(i)}_{\alpha \beta}+iM^{(i)}(x_{\mu})
(H^{(i)}_{m_i+\alpha, \beta}+H^{(i)}_{\alpha, m_i+ \beta}) 
+N^{(i)}(x_{\mu})H^{(i)}_{m_i+\alpha, m_i+ \beta} ,~~~~~~ \nonumber\\
G^{(i)}_{\alpha, m_i+\beta}
&=&L^{(i)}(x_{\mu}) H^{(i)}_{\alpha, m_i+ \beta}+iM^{(i)}(x_{\mu})
(H^{(i)}_{m_i+\alpha, m_i+ \beta}-H^{(i)}_{\alpha, \beta}) 
-N^{(i)}(x_{\mu})H^{(i)}_{m_i+\alpha, \beta} ,~~~~~~ \\
G^{(i)}_{m_i+\alpha, m_i+\beta}
&=&L^{(i)}(x_{\mu}) 
H^{(i)}_{m_i+\alpha, m_i+ \beta}-iM^{(i)}(x_{\mu})
(H^{(i)}_{\alpha, m_i+ \beta}+H^{(i)}_{m_i+\alpha, \beta}) 
+N^{(i)}(x_{\mu})H^{(i)}_{\alpha, \beta} ,~~~~~~\nonumber
\end{eqnarray}
where 
\begin{eqnarray}
L^{(i)}&=&-\frac{1}{A^{(i)}_{\mu}}
\frac{d}{dx_{\mu}}\tilde{X}_{\mu}
\frac{d}{dx_{\mu}}A^{(i)}_{\mu}
+\frac{1}{\tilde{X}_{\mu}}
\sum_{k,\ell=0}^{n-1}(-1)^{k+\ell}N_k N_{\ell}
x_{\mu}^{2(2n-k-\ell-2)}\nonumber\\
&+&\sum_{j=1(j\ne i)}^{N}
\frac{(-1)^{n+1}}{x_{\mu}^2-\xi_j^2} 
\frac{\Box^{(j)}K^{(j)}}{K^{(j)}}
-2\sum_{j=1}^{N}\frac{m_j x_{\mu}\tilde{X}_{\mu}}
{(x_{\mu}^2-\xi_j^2)A^{(i)}_{\mu}}
 \frac{d}{dx_{\mu}}A^{(i)}_{\mu}\\
&+& \frac{(-1)^{n+1}}{x_{\mu}^2} \frac{\Box^{(0)} K^{(0)}}{K^{(0)}}
- \frac{m^{(0)} \tilde{X}_{\mu}}{x_{\mu} A^{(i)}_{\mu}}
\frac{d}{dx_{\mu}} A_{\mu}^{(i)} \nonumber\\
&+&\frac{4\xi_i^2 \tilde{X}_{\mu}}{(x_{\mu}^2-\xi_i^2)^2}
+\frac{(-1)^{n+1}}{x_{\mu}^2-\xi_i^2}
(\Delta^{(i)}_L-2 \lambda^{(i)}),\nonumber\\
M^{(i)}&=&\frac{2\xi_i}{x_{\mu}^2-\xi_i^2}
\sum_{k=0}^{n-1}(-1)^k x_{\mu}^{2(n-k-1)}N_k,\\
N^{(i)}&=& 4\sum_{j=1}^{N}
\frac{\xi_i \xi_j \tilde{X}_{\mu}}{(x_{\mu}^2-\xi_i^2)
(x_{\mu}^2-\xi_j^2)}\nonumber
\end{eqnarray}
and
\begin{equation} \label{Gab}
G_{ab} = R(x_\mu) H^{(0)}_{ab},
\end{equation}
where
\begin{eqnarray}
R&=& -\frac{1}{B_{\mu}}\frac{d}{dx_{\mu}}
\tilde{X}_{\mu}\frac{d}{dx_{\mu}}B_{\mu}
+\frac{1}{\tilde{X}_{\mu}}
\sum_{k,\ell=0}^{n-1}(-1)^{k+\ell}N_k N_{\ell}
x_{\mu}^{2(2n-k-\ell-2)}\nonumber\\
&+&\sum_{i=1}^{N}\frac{(-1)^{n+1}}{x_{\mu}^2-\xi_i^2} 
\frac{\Box^{(i)}K^{(i)}}{K^{(i)}}
-2\sum_{i=1}^{N}\frac{m_i x_{\mu}\tilde{X}_{\mu}}
{(x_{\mu}^2-\xi_i^2)B_{\mu}}
 \frac{d}{dx_{\mu}}B_{\mu}\\
&-& \frac{m^{(0)} \tilde{X}_{\mu}}{x_{\mu} B_{\mu}}
 \frac{d}{dx_{\mu}}B_{\mu}
+\frac{(-1)^{n+1}}{x_{\mu}^2}
(\Delta^{(0)}_L-2 \lambda^{(0)}).\nonumber
\end{eqnarray}
In these expressions 
the function $\tilde{X}_{\mu}$ is defined by
\begin{equation}
\tilde{X}_{\mu}=\frac{1}{\displaystyle (x_{\mu})^{m^{(0)}-1} 
\prod_{i=1}^{N}(x_{\mu}^2-\xi_i^2)^{m_i}}
\left( d_{\mu} + \int \chi(x_{\mu})
x_{\mu}{}^{m^{(0)}-2} \prod_{i=1}^N
( x_{\mu}^2 - \xi_i^2)^{m_i} dx_{\mu}
\right).  
\end{equation}
The operator $\Delta^{(i)}_L$ is the gauge-covariant 
Lichnerowicz operator on $M^{(i)}$:
\begin{equation}
\Delta^{(i)}_L H^{(i)}_{\hat{\alpha} \hat{\beta}}=-
\sum_{\hat{\gamma}}\mathcal{D}^{(i)}_{\hat{\gamma}}
\mathcal{D}^{(i)}_{\hat{\gamma}}H^{(i)}_{\hat{\alpha} \hat{\beta}}
-2 \sum_{\hat{\gamma}, \hat{\delta}}
\tilde{R}^{(i)}_{\hat{\alpha} \hat{\gamma} \hat{\beta} \hat{\delta}}
H^{(i)}_{\hat{\gamma} \hat{\delta}}
+2 \lambda^{(i)} H^{(i)}_{\hat{\alpha} \hat{\beta}}
\end{equation}
and $\Box^{(i)}$ is the gauged scalar Laplacian on $M^{(i)}$:
\begin{equation}
\Box^{(i)}K^{(i)}=-\sum_{\hat{\alpha}}
\mathcal{D}^{(i)}_{\hat{\alpha}}
\bar{e}^{(i)}_{\hat{\alpha}}K^{(i)}.
\end{equation}
The Lichnerowicz operator $\Delta^{(0)}_L$ 
on $M^{(0)}$ is defined by
\begin{equation}
\Delta^{(0)}_L H^{(0)}_{ab}=-
\sum_{d}D^{(0)}_{d} D^{(0)}_{d}H^{(0)}_{ab}
-2 \sum_{de}
\tilde{R}^{(i)}_{adbe}H^{(0)}_{de}
+2 \lambda^{(0)} H^{(0)}_{ab}
\end{equation}
and $\Box^{(0)}$ is the scalar Laplacian on $M^{(0)}$:
\begin{equation}
\Box^{(0)} K^{(0)} = - \sum_a D_a^{(0)} \tilde{e}_a K^{(0)}.
\end{equation}
The cosmological constants $\lambda^{(i)}$ and 
$\lambda^{(0)}$ are given by \eqref{cosmi} and \eqref{al0}.

The solutions of \eqref{UG} are
\begin{equation} \label{solG}
G^{(i)}_{\hat{\alpha} \hat{\beta}}
=\sum_{k=0}^{n-2} b^{(i)}_k x_{\mu}^{2k} 
H^{(i)}_{\hat{\alpha} \hat{\beta}},~~~
G_{ab}=\sum_{k=0}^{n-2} b_k x_{\mu}^{2k} 
H^{(0)}_{ab},
\end{equation}
where $b^{(i)}_k$ and $b_k$ are arbitrary constants. 
This is shown by employing the identity
\begin{equation}
\sum_{\mu=1}^{n}\frac{x_{\mu}^{2k}}{U_{\mu}}=0~~~~~
\mbox{for}~~ k=0, \cdots, n-2,
\end{equation}
which also played crucial roles in \cite{OY2,FKK,HHOY}.

Now
we further simplify the perturbation equation
with help of the K\"{a}hler condition.
We can take the tensor 
$H^{(i)}_{\hat{\alpha} \hat{\beta}}$ 
\begin{equation}
\bullet~\mbox{hermitian}~: ~~H^{(i)}_{\alpha \beta}
=H^{(i)}_{m_i+\alpha, m_i+ \beta},~~
H^{(i)}_{\alpha, m_i+ \beta}
=-H^{(i)}_{m_i+\alpha, \beta}~~~~~~~~~~~~~~~~~~~~~
\end{equation}
~~~~~~~or
\begin{equation}
\bullet~\mbox{anti-hermitian}~: ~~H^{(i)}_{\alpha \beta}
=-H^{(i)}_{m_i+\alpha, m_i+ \beta},~
H^{(i)}_{\alpha, m_i+ \beta}
=H^{(i)}_{m_i+\alpha, \beta}.~~~~~~~~~~~~~~~
\end{equation}
This means that $H^{(i)}_{\hat{\alpha} \hat{\beta}}$ 
is an eigenfunction of the linear map $\mathcal{J}^{(i)}$
\cite{KLR}
\begin{equation}
(\mathcal{J}^{(i)} H^{(i)})_{\hat{\alpha} \hat{\beta}}
=J^{(i)}_{\hat{\alpha}}{}^{\hat{\gamma}}
J^{(i)}_{\hat{\beta}}{}^{\hat{\delta}}
H^{(i)}_{\hat{\gamma} \hat{\delta}},
\end{equation}
where $J^{(i)}$ is given by \eqref{Kahler}. Indeed, the map has the 
eigenvalues $\pm 1$ and their eigenfunctions are
hermitian or anti-hermitian, respectively.
Note that the condition \eqref{DH1} is now a consequence 
of the transverse condition \eqref{DH2}. Furthermore the K\"ahler
condition \eqref{wKc} leads to the commutativity 
$[ \Delta_L^{(i)}, \mathcal{J}^{(i)} ]=0$ 
on the rank-2 tensor space.
Thus one can choose the simultaneous eigenfunction 
$H^{(i)}_{\hat{\alpha} \hat{\beta}}$ of the operators
 $\Delta_L^{(i)}$ and $\mathcal{J}^{(i)}$. 
It is also assumed 
that $K^{(i)}$, $K^{(0)}$ and $H^{(0)}_{ab}$ are eigenfunctions
of $\Box^{(i)}$, $\Box^{(0)}$ and $\Delta_L^{(0)}$ , and 
$\{H^{(i)}_{\hat{\alpha} \hat{\beta}}, H^{(0)}_{ab} \}$ 
satisfies the traceless and transverse conditions.
\begin{enumerate}
\item[(a)] hermitian:~~Combining \eqref{exG} \eqref{Gab} 
with \eqref{solG} we have
\begin{equation} \label{LNR}
L^{(i)}+N^{(i)}=\sum_{k=0}^{n-2} b_k^{(i)} x_{\mu}^{2k},~~~
R=\sum_{k=0}^{n-2} b_k x_{\mu}^{2k}.
\end{equation}
\item[(b)] anti-hermitian :~~We have
\begin{eqnarray}
\left(L^{(i)}-N^{(i)}-\sum_{k=0}^{n-2} b_k^{(i)} 
x_{\mu}^{2k} \right) H^{(i)}_{\alpha \beta}
+2 i M^{(i)}H^{(i)}_{\alpha, m_i+ \beta}&=&0,\\
-2 i M^{(i)}H^{(i)}_{\alpha \beta}+
\left(L^{(i)}-N^{(i)}-\sum_{k=0}^{n-2} b_k^{(i)} 
x_{\mu}^{2k} \right)H^{(i)}_{\alpha,m_i+ \beta}&=&0.
\end{eqnarray}
In order to allow the non-zero eigenfunction 
$H^{(i)}_{\hat{\alpha} \hat{\beta}}$,
it is necessary to satisfy the condition
\begin{equation}
L^{(i)}-N^{(i)}+2 \epsilon M^{(i)}
-\sum_{k=0}^{n-2} b_k^{(i)} x_{\mu}^{2k} =0
\end{equation}
with $\epsilon=\pm 1$. We also obtain 
the second equation in \eqref{LNR}.
\end{enumerate}
Thus we have demonstrated that the equation \eqref{linEin} 
in the generalized Kerr-NUT-de Sitter
background allows a separation of variables \eqref{sephi} \eqref{seph0}
when the functions $A^{(i)}_{\mu}$ and $B_{\mu}$
satisfy the ordinary second order differential equations:
\begin{equation} \label{ODE1}
-\frac{d}{dx_{\mu}}\tilde{X}_{\mu} 
\frac{d}{dx_{\mu}}A^{(i)}_{\mu}
-\left( 2\sum_{j=1}^N \frac{m_j x_{\mu} 
\tilde{X}_{\mu}}{x_{\mu}^2-\xi_j^2} 
+ \frac{m^{(0)} \tilde{X}_{\mu}}{x_{\mu}} \right)
\frac{d}{dx_{\mu}} A_{\mu}^{(i)}
+V_{\mu}^{(i)} A_{\mu}^{(i)}=0,
\end{equation}
\begin{eqnarray}
V_{\mu}^{(i)}&=&\frac{1}{\tilde{X}_{\mu}}
\sum_{k,\ell=0}^{n-1}(-1)^{k+\ell} 
N_k N_{\ell} x_{\mu}^{2(2n-k-\ell-2)}
+\sum_{j=1(j \ne i)}^{N}
\frac{(-1)^{n+1} E^{(j)}_s}{x_{\mu}^2-\xi_j^2}
+(-1)^{n+1} \frac{E_s^{(0)}}{x_{\mu}^2}
\nonumber\\
&+&\frac{4 \xi_i^2 \tilde{X}_{\mu}}{(x_{\mu}^2-\xi_i^2)^2}
+\frac{(-1)^{n+1}(E^{(i)}_t-2 \lambda^{(i)})}{x_{\mu}^2-\xi_i^2}
-\sum_{k=0}^{n-2} b^{(i)}_k x_{\mu}^{2k} \\
&+&4\sigma \sum_{j=1}^{N} 
\frac{\xi_i \xi_j \tilde{X}_{\mu}}
{(x_{\mu}^2-\xi_i^2)(x_{\mu}^2-\xi_j^2)}
+\frac{\epsilon(1-\sigma)2\xi_i}
{x_{\mu}^2-\xi_i^2}
\sum_{k=0}^{n-1} (-1)^k x_{\mu}^{2(n-k-1)} N_k\nonumber
\end{eqnarray}
and
\begin{equation} \label{ODE2}
-\frac{d}{dx_{\mu}}\tilde{X}_{\mu} \frac{d}{dx_{\mu}}B_{\mu}
-\left( 2 \sum_{j=1}^N \frac{m_j x_{\mu} 
\tilde{X}_{\mu}}{x_{\mu}^2-\xi_j^2}
+\frac{m^{(0)} \tilde{X}_{\mu}}{x_{\mu}} \right) 
\frac{d}{dx_{\mu}}B_{\mu}+V_{\mu}^{(0)} B_{\mu}=0,
\end{equation}
\begin{eqnarray}
V_{\mu}^{(0)}&=&\frac{1}{\tilde{X}_{\mu}}
\sum_{k,\ell=0}^{n-1}(-1)^{k+\ell} 
N_k N_{\ell} x_{\mu}^{2(2n-k-\ell-2)}
+\sum_{j=1}^{N}\frac{(-1)^{n+1} E^{(j)}_s}
{x_{\mu}^2-\xi_j^2}\nonumber\\
&+&\frac{(-1)^{n+1}(E^{(0)}_t-2 \lambda^{(0)})}{x_{\mu}^2}
-\sum_{k=0}^{n-2} b_k x_{\mu}^{2k}, \nonumber
\end{eqnarray}
where according to hermitian and anti-hermitian 
the symbol $\sigma$ takes the values $\pm 1$, and
$E^{(i)}_s, E^{(i)}_t, E_s^{(0)}$ and $ E^{(0)}_t$ represent 
the eigenvalues of $\Box^{(i)}, \Delta_L^{(i)}$, $\Box^{(0)}$ and
$\Delta_L^{(0)}$, respectively.

\subsection{Special type}
The procedure is completely parallel to the case of general type. 
The equation \eqref{linEin} allows
a separation of variables for the tensor components
\begin{equation}
h_{(\hat{\alpha},i),(\hat{\beta},i)}
=\left( \prod_{\mu=1}^{n} \hat{A}^{(i)}_{\mu}(x_{\mu})
\prod_{k=0}^{n} e^{i N_k \psi_k} \right)
H^{(i)}_{\hat{\alpha} \hat{\beta}}(y_I^{(i)}) 
\prod_{\stackrel{\scriptstyle j=1}{ (j \ne i)}}^{N} 
K^{(j)}(y_J^{(j)}),
\end{equation}
where $ H^{(i)}_{\hat{\alpha} \hat{\beta}}(y_I^{(i)})$ 
is the simultaneous eigenfunction of the
operators $\Delta_L^{(i)}$ and 
$\mathcal{J}^{(i)}$ satisfying the traceless 
and transverse conditions,
and $K^{(i)}$ the eigenfunction of  $\Box^{(i)}$. 
The functions $\hat{A}^{(i)}_{\mu}$ 
satisfy the ordinary second order differential equations:
\begin{equation} \label{ODE3}
-\frac{d}{dx_{\mu}}\tilde{X}_{\mu} 
\frac{d}{dx_{\mu}}\hat{A}^{(i)}_{\mu}
-\left( 2\sum_{j=1}^N \frac{m_j x_{\mu} 
\tilde{X}_{\mu}}{x_{\mu}^2-\xi_j^2}
+\frac{\tilde{X}_{\mu}}{x_{\mu}}
\right) \frac{d}{dx_{\mu}}\hat{A}_{\mu}^{(i)}
+\hat{V}_{\mu}^{(i)} \hat{A}_{\mu}^{(i)}=0,
\end{equation}
\begin{eqnarray}
\hat{V}_{\mu}^{(i)}&=&\frac{1}{\tilde{X}_{\mu}}
\sum_{k,\ell=0}^{n}(-1)^{k+\ell} N_k N_{\ell} 
x_{\mu}^{2(2n-k-\ell-2)}
+\frac{(-1)^{n+1} N_n^2}{c x_{\mu}^2}
+\sum_{j=1(j \ne i)}^{N}
\frac{(-1)^{n+1} E^{(j)}_s}{x_{\mu}^2-\xi_j^2}\nonumber\\
&+&\frac{4c}{\xi_i^2} \frac{(-1)^{n+1}}{x_{\mu}^2}
+ \frac{4 \xi_i^2 \tilde{X}_{\mu}}{(x_{\mu}^2-\xi_i^2)^2}
+\frac{(-1)^{n+1}(E^{(i)}_t-2 \lambda^{(i)})}{x_{\mu}^2-\xi_i^2}
-\sum_{k=0}^{n-2} \hat{b}^{(i)}_k x_{\mu}^{2k} \\
&+& 4\sigma \sum_{j=1}^{N} 
\frac{\xi_i \xi_j \tilde{X}_{\mu}}
{(x_{\mu}^2-\xi_i^2)(x_{\mu}^2-\xi_j^2)}
+\frac{\epsilon(1-\sigma)2\xi_i}{x_{\mu}^2-\xi_i^2}
\sum_{k=0}^{n} (-1)^k x_{\mu}^{2(n-k-1)} N_k\nonumber\\
&+&4 \sigma \sum_{j=1}^{N} 
\frac{(-1)^{n+1} c}{\xi_i \xi_j x_{\mu}^2}-
\frac{(-1)^{n+1}2 \epsilon (1-\sigma) N_n}
{\xi_i x_{\mu}^2},\nonumber
\end{eqnarray}
where $\hat{b}^{(i)}_k$ are arbitrary constants and
\beq
\tilde{X}_{\mu}= 
\prod_{i=1}^N ( x_{\mu}^2 - \xi_i^2)^{-m_i}
\left( d_{\mu} + \int \chi(x_{\mu}) x_{\mu}^{-1}
\prod_{i=1}^N (x_{\mu}^2 - \xi_i^2)^{m_i} d x_{\mu} \right).
\eeq


\section{Summary and discussion}

In this paper we have studied the separability 
of the gravitational perturbations
in the generalized Kerr-NUT-de Sitter spacetimes.
We found that tensor type perturbations admit 
the separation of variables and the equations of motion
reduce to a set of ordinary second order differential equations.
It seems to be sure that the separability 
is deeply related to the existence of the conformal 
Killing-Yano tensor like the cases of geodesic equation, 
Klein-Gordon equation and
Dirac equation. However,
the geometrical origin still remains veiled. 
It is important to clarify
why the separability works well, and also important
to study whether the symmetry connected with CKY tensor
\cite{KaTra}
enables the separation for more general perturbations.

Our results can be used for the study of the stability of 
higher-dimensional Kerr-de Sitter
black holes with partially equal angular momenta.
The expressions for the metrics given in \cite{GLPP1,GLPP2} 
are rather
complicated and hence they are not so convenient 
for the perturbations.
We found the explicit coordinate transformations 
from the black holes to the
generalized Kerr-NUT-de Sitter spacetimes. 
Thus our formulation applies to
the Kerr-de Sitter black holes with such 
angular momenta\footnote{Unfortunately, 
our formulation cannot apply to 
the non-degenerate angular momenta. This is a future problem.}. 
In order to investigate the problem of
the stability  we must specify the several quantities
in the equations \eqref{ODE1}, \eqref{ODE2} and \eqref{ODE3}.
First we need to 
know the eigenvalues of the (gauged) scalar Laplacian
and (gauge-covariant) 
Lichnerowicz operator
on K\"ahler-Einstein manifolds and Einstein manifolds. 
Fortunately, 
in the case of the Kerr-de Sitter
black holes these Einstein manifolds are 
complex projective spaces and standard spheres, 
on which the eigenvalues are well known 
(see for example \cite{GP,IT,war,pop,PS,HMP,bou1,KLR,bou2,bou3}). 
Next, we must
determine the ranges of the Killing coordinates. 
If we choose them suitably,
then the charge given by \eqref{cni} 
is proportional to an integer, 
which is interpreted  as the
first Chern number of the line bundle over 
the K\"ahler-Einstein manifold. The equations 
\eqref{ODE1}, \eqref{ODE2} and \eqref{ODE3} are 
Fuchs type differential equations, and the
boundary condition to the solutions will be given  
according to the analysis in \cite{KLR,KKZ}. 
We hope to report our stability analysis of the 
higher-dimensional Kerr-de Sitter black holes 
in a separated paper.

\vspace{5mm}

\noindent
{\bf{Acknowledgements}}

\vspace{3mm}

We would like to thank Tsuyoshi Houri  
and Hideki Ishihara for discussions.
The work of YY is supported by the Grant-in Aid for Scientific
Research (No. 19540304 and No. 19540098)
from Japan Ministry of Education. 
The work of TO is supported by the Grant-in Aid for Scientific
Research (No. 19540304 and No. 20540278)
from Japan Ministry of Education.


\appendix

\section{Spin connections and curvature $2$-forms
for the general type}

In this appendix, we give explicit forms of the spin connections
and curvature $2$-forms for the metric \eqref{gCKY} of the general type.

For the metric \eqref{gCKY},
we introduce the following orthonormal frame 
$\{e^A\}=\{e^{\mu}, e^{n+\mu}, e^{a}, e^{\hat{\alpha}}_{(i)} \}$:
\begin{eqnarray}
e^{\mu} &=& \frac{d x_{\mu}}{\sqrt{P_{\mu}}}, \qquad
e^{n+\mu} = \sqrt{P_{\mu}} 
\left( \sum_{k=0}^{n-1} 
\sigma_k( \tilde{x}_{\mu}) \theta_k \right), \nonumber\\
e^{a} &=& \sqrt{\sigma_n} \tilde{e}^{a}, \qquad
e^{\hat{\alpha}}_{(i)} = \left( \prod_{\mu=1}^n 
( x_{\mu}^2-\xi_i^2 ) \right)^{1/2} 
\tilde{e}^{\hat{\alpha}}_{(i)}.
\end{eqnarray}
Here, $\{ \tilde{e}^{a} \}_{a=1,2,\dotsc, m^{(0)}}$ 
is an orthonormal frame
of a Riemann manifold $(M^{(0)},g^{(0)})$,
and $\{\tilde{e}^{\hat{\alpha}}_{(i)} \}=
\{\tilde{e}^{\alpha}_{(i)}, 
\tilde{e}^{m_i+\alpha}_{(i)} \}_{\alpha=1,2,\dotsc, m_i}$ 
are orthonormal frames of 
K\"ahler manifolds $(M^{(i)}, g^{(i)}, J^{(i)}, \omega^{(i)})$
$(i=1, \cdots, N)$
such that the K\"ahler structure is of the form
\begin{eqnarray} \label{Kahler}
&\bullet& \mbox{metric~:}~~~~~~~~~g^{(i)}=
\sum_{\alpha=1}^{m_i}(\tilde{e}^{\alpha}_{(i)}
\otimes \tilde{e}^{\alpha}_{(i)}+
\tilde{e}^{m_i+\alpha}_{(i)} \otimes 
\tilde{e}^{m_i+\alpha}_{(i)} ),\nonumber\\
&\bullet& \mbox{complex structure~:}~~~~~~~~~ 
J^{(i)}(\tilde{e}^{\alpha}_{(i)}) 
= -\tilde{e}^{m_i+\alpha}_{(i)},~~~
J^{(i)}(\tilde{e}^{m_i+\alpha}_{(i)}) 
= \tilde{e}^{\alpha}_{(i)},\\
&\bullet& \mbox{K\"ahler form~:}~~~~~~~~~ 
\omega^{(i)}= \sum_{\alpha=1}^{m_i} 
\tilde{e}^{\alpha}_{(i)} \wedge
\tilde{e}^{m_i+\alpha}_{(i)}.\nonumber
\end{eqnarray}
The dual vector fields defined by 
$e^{A}(e_{B})=\delta^A_B$ are written as
\begin{eqnarray} \label{gdualv}
e_{\mu} &=& \sqrt{P_{\mu}}
\frac{\partial}{\partial x_{\mu}},\nonumber\\
e_{n+\mu}&=& \frac{1}{U_{\mu} \sqrt{P_{\mu}}} 
\sum_{k=0}^{n-1}(-1)^k x_{\mu}^{2(n-k-1)} 
\frac{\partial}{\partial \psi_k}, \nonumber\\
e_{a}&=& \frac{1}{\sqrt{\sigma_n}} \tilde{e}_{a},\\
e_{\hat{\alpha}}^{(i)}&=& 
\left( \prod_{\mu=1}^n (x_{\mu}^2-\xi_i^2) 
\right)^{-1/2} 
\bar{e}_{\hat{\alpha}}^{(i)},\nonumber
\end{eqnarray}
where 
\begin{equation} \label{gbei}
\bar{e}_{\hat{\alpha}}^{(i)}
=\tilde{e}_{\hat{\alpha}}^{(i)}
+2 A^{(i)}_{\hat{\alpha}} \sum_{k=0}^{n-1}(-1)^{n+k} 
\xi_i^{2(n-k)-1} \frac{\partial}{\partial \psi_k}.
\end{equation}
The vector fields
$\tilde{e}_{a}$ and  $\tilde{e}_{\hat{\alpha}}^{(i)}$ 
are the dual vector fields to the 1-forms
$\tilde{e}^{a}$ and $\tilde{e}^{\hat{\alpha}}_{(i)}$, 
respectively. 
The $A^{(i)}_{\hat{\alpha}}$
represents the component
of the 1-form, $A^{(i)}=A^{(i)}_{\hat{\alpha}} 
\tilde{e}^{\hat{\alpha}}_{(i)}$.

The connection $1$-forms $\omega_{AB}=-\omega_{BA}$, which obey 
the first structure equation
\begin{equation}
d e^A + \omega^A{}_B \wedge e^B = 0,
\end{equation}
are determined as follows:
\begin{equation}
\begin{split}
\omega_{\mu \nu} &= 
\frac{(1-\delta_{\mu \nu})}{ x_{\mu}^2 - x_{\nu}^2}
\left( - x_{\nu} \sqrt{P_{\nu}} e^{\mu} 
- x_{\mu} \sqrt{P_{\mu}} e^{\nu} \right), \cr
\omega_{\mu, n+\nu}
&= \delta_{\mu\nu}
\left[ - \frac{\partial}{\partial x_{\mu}}( \sqrt{P}_{\mu})
e^{n+\mu} + \sum_{\rho=1}^n 
\frac{(1-\delta_{\mu \rho}) x_{\mu} \sqrt{P_{\rho}}}
{x_{\mu}^2-x_{\rho}^2}
e^{n+\rho} \right] \cr
& + \frac{(1-\delta_{\mu\nu})}{x_{\mu}^2-x_{\nu}^2}
\left( x_{\mu} \sqrt{P_{\nu}} e^{n+\mu} - x_{\mu} 
\sqrt{P_{\mu}} e^{n+\nu} \right),~~~(\mbox{no sum}), \cr
\omega_{n+\mu, n+\nu}
&= \frac{(1-\delta_{\mu\nu})}{x_{\mu}^2 - x_{\nu}^2}
\left( - x_{\nu} \sqrt{P_{\mu}} e^{\nu} - x_{\mu} 
\sqrt{P_{\nu}} e^{\mu} \right),
\end{split}
\end{equation}
\begin{align}
\omega_{\mu, (\alpha,i)}
&= - \frac{x_{\mu} \sqrt{P_{\mu}}}{x_{\mu}^2-\xi_i^2} 
e^{\alpha}_{(i)},&
\omega_{\mu, (m_i+\alpha,i)}
&= - \frac{x_{\mu} \sqrt{P_{\mu}}}{x_{\mu}^2 - \xi_i^2}
e^{m_i+\alpha}_{(i)}, \cr
\omega_{n+\mu, (\alpha,i)}
&= \frac{\xi_i \sqrt{P_{\mu}}}{x_{\mu}^2 - \xi_i^2}
e^{m_i+\alpha}_{(i)},&
\omega_{n+\mu, (m_i+\alpha,i)}
&= - \frac{\xi \sqrt{P_{\mu}} }{x_{\mu}^2 - \xi_i^2}
e^{\alpha}_{(i)},
\end{align}
\begin{equation}
\omega_{\mu a}= - \frac{\sqrt{P_{\mu}}}{x_{\mu}} 
e^{a}, \qquad
\omega_{n+\mu,a} =0,
\end{equation}
\begin{equation}
\begin{split}
\omega_{(\alpha,i), ( \beta,j)} &= \delta_{ij} 
\tilde{\omega}^{(i)}_{\alpha \beta}, \cr
\omega_{(\alpha,i), (m_j+\beta,j)}
&= \delta_{ij} \left(
\tilde{\omega}^{(i)}_{\alpha, m_i+\beta}
- \delta_{\alpha \beta} \sum_{\mu=1}^n 
\frac{\xi_i \sqrt{P_{\mu}}}{x_{\mu}^2 - \xi_i^2} 
e^{n+\mu} \right), \cr
\omega_{(m_i+\alpha, i), (m_j+\beta,j)} 
&= \delta_{ij} \tilde{\omega}^{(i)}_{m_i+\alpha, m_i+\beta},
\end{split}
\end{equation}
\begin{equation}
\omega_{(\alpha,i),b}=\omega_{(m_i+\alpha,i),b}=0,
\end{equation}
\begin{equation}
\omega_{ab} = \tilde{\omega}_{ab}.
\end{equation}
Here $\tilde{\omega}_{ab}$ and
$\tilde{\omega}^{(i)}_{\hat{\alpha} \hat{\beta}}$
are connection $1$-forms
for the metrics of $g^{(0)}$ and $g^{(i)}$, respectively.
Since $g^{(i)}$ is a K\"{a}hler metric, 
the $1$-forms 
$\tilde{\omega}^{(i)}_{\hat{\alpha} \hat{\beta}}$ 
obey the following conditions:
\begin{equation} \label{wKc}
\tilde{\omega}^{(i)}_{m_i+\alpha, \beta} 
= - \tilde{\omega}^{(i)}_{\alpha,  m_i+\beta}, 
\qquad
\tilde{\omega}^{(i)}_{m_i+\alpha, m_i+\beta} 
= \tilde{\omega}^{(i)}_{\alpha \beta}.
\end{equation}
To represent the components of 
the curvature $2$-forms $R^A{}_{B}$ 
it is convenient to introduce the following functions:
\begin{equation}
P_T^{[k]}(t):= \sum_{\mu=1}^n 
\frac{P_{\mu}}{(x_{\mu}^2 - t)^k},~~~P_T:= P_T^{[0]}(t) 
= \sum_{\mu=1}^n P_{\mu}.
\end{equation}
We also use the 2-form
\begin{equation}
W^{(i)}:= \sum_{\alpha=1}^{m_i} e^{\alpha}_{(i)} 
\wedge e^{m_i+\alpha}_{(i)}.
\end{equation}
From the second structure equation
\begin{equation}
R^{A}{}_{B} = d \omega^{A}{}_{B} 
+ \omega^{A}{}_{C} \wedge \omega^C{}_B,
\end{equation}
we find the explicit form of the curvature
$2$-forms. For $(\mu \neq \nu)$, we have
\begin{equation}
\begin{split}
R_{\mu \nu}&=
 - \frac{1}{2(x_{\mu}^2 - x_{\nu}^2)}
\left( x_{\mu} \frac{\partial P_T}{\partial x_{\mu}} 
- x_{\nu} \frac{\partial P_T}{\partial x_{\nu}} \right) 
e^{\mu} \wedge e^{\nu} \cr
& \ \ \ - \frac{1}{2(x_{\mu}^2 - x_{\nu}^2)}
\left( x_{\nu} \frac{\partial P_T}{\partial x_{\mu}}
- x_{\mu} \frac{\partial P_T}{\partial x_{\nu}} \right)
e^{n+\mu} \wedge e^{n+\nu},
\end{split}
\end{equation}
\begin{equation}
\begin{split}
R_{\mu, n+\mu}
&= - \frac{1}{2} \frac{\partial^2 P_T}{\partial x_{\mu}^2} 
e^{\mu} \wedge e^{n+\mu} \cr
& \ \ \ + \sum_{\rho \neq \mu}
\frac{1}{x_{\mu}^2 - x_{\rho}^2}
\left( x_{\mu} \frac{\partial P_T}{\partial x_{\rho}}
- x_{\rho} \frac{\partial P_T}{\partial x_{\mu}} \right) 
e^{\rho} \wedge e^{n+\rho} \cr
& \ \ \ 
-\sum_{i=1}^N \xi_i 
\frac{\partial P_T^{[1]}(\xi_i^2)}{\partial x_{\mu}}
W^{(i)}, \qquad (\mbox{no sum}),
\end{split}
\end{equation}
\begin{equation}
\begin{split}
R_{\mu, n+\nu}
&=- \frac{1}{2(x_{\mu}^2 - x_{\nu}^2)}
\left( x_{\mu} \frac{\partial P_T}{\partial x_{\mu}}
- x_{\nu} \frac{\partial P_T}{\partial x_{\nu}} \right)
e^{\mu} \wedge e^{n+\nu} \cr
& \ \ \ 
+ \frac{1}{2(x_{\mu}^2 - x_{\nu}^2)}
\left( x_{\mu} \frac{\partial P_T}{\partial x_{\nu}}
- x_{\nu} \frac{\partial P_T}{\partial x_{\mu}} \right)
e^{\nu} \wedge e^{n+\mu},
\end{split}
\end{equation}
\begin{equation}
\begin{split}
R_{n+\mu, n+\nu}
&= - \frac{1}{ 2(x_{\mu}^2 - x_{\nu}^2 ) }
\left(
x_{\nu} \frac{\partial P_T}{\partial x_{\mu}}
- x_{\mu} \frac{\partial P_T}{\partial x_{\nu}}
\right) e^{\mu} \wedge e^{\nu} \cr
& \ \ \ 
- \frac{1}{ 2(x_{\mu}^2 - x_{\nu}^2) }
\left(
x_{\mu} \frac{\partial P_T}{\partial x_{\mu}}
- x_{\nu} \frac{\partial P_T}{\partial x_{\nu}}
\right) e^{n+\mu} \wedge e^{n+\nu},
\end{split}
\end{equation}
\begin{equation}
R_{\mu, (\alpha,i)}
= - \left[ \left( 1 + \frac{1}{2} x_{\mu} 
\frac{\partial}{\partial x_{\mu}} \right)
P_T^{[1]}(\xi_i^2) \right] e^{\mu} \wedge e^{\alpha}_{(i)}
 - \frac{1}{2} \xi_i 
\frac{\partial P_T^{[1]}(\xi_i^2)}{\partial x_{\mu}}
e^{n+\mu} \wedge e^{m_i+\alpha}_{(i)},
\end{equation}
\begin{equation}
R_{\mu, (m_i+\alpha,i)}
= -\left[ \left( 1 + \frac{1}{2} x_{\mu} 
\frac{\partial}{\partial x_{\mu}} \right)
P_T^{[1]}(\xi_i^2) \right] e^{\mu} \wedge e^{m_i+\alpha}_{(i)}
 + \frac{1}{2} \xi_i 
\frac{\partial P_T^{[1]}(\xi_i^2)}{\partial x_{\mu}}
e^{n+\mu} \wedge e^{\alpha}_{(i)},
\end{equation}
\begin{equation}
R_{n+\mu, (\alpha,i)}
=  \frac{1}{2} \xi_i \frac{\partial P_T^{[1]}(\xi_i^2)}
{\partial x_{\mu}}
e^{\mu} \wedge e^{m_i+\alpha}_{(i)}
-\left[ \left( 1 + \frac{1}{2} x_{\mu} 
\frac{\partial}{\partial x_{\mu}} \right)
P_T^{[1]}(\xi_i^2) \right] 
e^{n+\mu} \wedge e^{\alpha}_{(i)},
\end{equation}
\begin{equation}
R_{n+\mu, (m_i+\alpha,i)}
= -\frac{1}{2} \xi_i 
\frac{\partial P_T^{[1]}(\xi_i^2)}{\partial x_{\mu}}
e^{\mu} \wedge e^{\alpha}_{(i)}
-\left[ \left( 1 + \frac{1}{2} x_{\mu} 
\frac{\partial}{\partial x_{\mu}} \right)
P_T^{[1]}(\xi_i^2) \right] 
e^{n+\mu} \wedge e^{m_i+\alpha}_{(i)},
\end{equation}
\begin{equation}
\begin{split}
R_{\mu,a} 
=-\left[ \left( 1 + \frac{1}{2} x_{\mu} 
\frac{\partial}{\partial x_{\mu}} \right)
P_T^{[1]}(0) \right] 
e^{\mu} \wedge e^{a},
\end{split}
\end{equation}
\begin{equation}
R_{n+\mu,a}
=-\left[ \left( 1 + \frac{1}{2} x_{\mu} 
\frac{\partial}{\partial x_{\mu}} \right)
P_T^{[1]}(0) \right] 
e^{n+\mu} \wedge e^{a}.
\end{equation}

For $\alpha \neq \beta$ we have
\begin{equation}
\begin{split}
R_{(\alpha,i),(\beta,i)}
&= 
\tilde{R}^{(i)}_{\alpha \beta} - 
\left( P_T^{[1]}(\xi_i^2)+ \xi_i^2 P_T^{[2]}(\xi_i^2) \right)
e^{\alpha}_{(i)} \wedge e^{\beta}_{(i)} \cr
& \qquad - \xi_i^2 P_T^{[2]}(\xi_i^2) \, e^{m_i+\alpha}_{(i)}
\wedge e^{m_i+\beta}_{(i)}, 
\end{split}
\end{equation}
\begin{equation}
\begin{split}
R_{(\alpha,i), (m_i+\alpha,i)}
&= \tilde{R}^{(i)}_{\alpha, m_i+\alpha}
- \xi_i \sum_{\mu=1}^n
\frac{\partial P_T^{[1]}(\xi_i^2) }{\partial x_{\mu} }\, 
e^{\mu} \wedge e^{n+\mu} \cr
& - 2 \xi_i^2 P_T^{[2]}(\xi_i^2) \, W^{(i)}
- 2 \sum_{\stackrel{\scriptstyle j=1}{(j \neq i)}}^N
\frac{\xi_i \xi_j}{\xi_i^2 - \xi_j^2}
\left( P_T^{[1]}(\xi_i^2) - P_T^{[1]}(\xi_j^2) \right)
W^{(k)} \cr
& - \left( P_T^{[1]}(\xi_i^2)
+ 2 \xi_i^2 P_T^{[2]}(\xi_i^2) \right)
e^{\alpha}_{(i)} \wedge e^{m_i+\alpha}_{(i)}, \qquad
(\mbox{no sum}),
\end{split}
\end{equation}
\begin{equation}
\begin{split}
R_{(\alpha,i),(m_i+\beta,i)}
&= 
\tilde{R}^{(i)}_{\alpha, m_i+\beta} 
+ \xi_i^2 P_T^{[2]}(\xi_i^2) \, e^{m_i+\alpha}_{(i)}
\wedge e^{\beta}_{(i)} \cr
& \qquad - 
\left( P_T^{[1]}(\xi_i^2)+ \xi_i^2 P_T^{[2]}(\xi_i^2) \right)
e^{\alpha}_{(i)} \wedge e^{m_i+\beta}_{(i)}, 
\end{split}
\end{equation}
\begin{equation}
\begin{split}
R_{(m_i+\alpha,i),(m_i+\beta,i)}
&= 
\tilde{R}^{(i)}_{m_i+\alpha, m_i+\beta} - 
\xi_i^2 P_T^{[2]}(\xi_i^2) \, e^{\alpha}_{(i)}
\wedge e^{\beta}_{(i)} \cr
& \qquad - 
\left( P_T^{[1]}(\xi_i^2)+ \xi_i^2 P_T^{[2]}(\xi_i^2) \right)
e^{m_i+\alpha}_{(i)} \wedge e^{m_i+\beta}_{(i)}.
\end{split}
\end{equation}

For general $\alpha$, $\beta$ and $i \neq j$ we have
\begin{equation}
\begin{split}
R_{(\alpha,i),(\beta,j)}
&= 
-\frac{1}{\xi_i^2-\xi_j^2} 
\left( \xi_i^2 P_T^{[1]}(\xi_i^2)
- \xi_j^2 P_T^{[1]}(\xi_j^2)  \right)
e^{\alpha}_{(i)} \wedge e^{\beta}_{(j)} \cr
& \qquad - \frac{\xi_i \xi_j}{\xi_i^2-\xi_j^2}
(P_T^{[1]}(\xi_i^2)-P_T^{[1]}(\xi_j^2)) \, 
e^{m_i+\alpha}_{(i)}
\wedge e^{m_i+\beta}_{(j)}, 
\end{split}
\end{equation}
\begin{equation}
\begin{split}
R_{(\alpha,i), (m_i+\alpha,j)}
&= 
-\frac{1}{\xi_i^2-\xi_j^2} 
\left( \xi_i^2 P_T^{[1]}(\xi_i^2)
- \xi_j^2 P_T^{[1]}(\xi_j^2)  \right)
e^{\alpha}_{(i)} \wedge e^{m_j+\beta}_{(j)} \cr
& \qquad - \frac{\xi_i \xi_j}{\xi_i^2-\xi_j^2}
(P_T^{[1]}(\xi_i^2)
-P_T^{[1]}(\xi_j^2)) \, e^{m_i+\alpha}_{(i)}
\wedge e^{\beta}_{(j)}, 
\end{split}
\end{equation}
\begin{equation}
\begin{split}
R_{(m_i+\alpha,i),(m_j+\beta,j)}
&= 
-\frac{1}{\xi_i^2-\xi_j^2} 
\left( \xi_i^2 P_T^{[1]}(\xi_i^2)
-\xi_j^2 P_T^{[1]}(\xi_j^2)  \right)
e^{m_i+\alpha}_{(i)} 
\wedge e^{m_j+\beta}_{(j)} \cr
& \qquad - \frac{\xi_i \xi_j}{\xi_i^2-\xi_j^2}
(P_T^{[1]}(\xi_i^2)
-P_T^{[1]}(\xi_j^2)) \, e^{\alpha}_{(i)}
\wedge e^{\beta}_{(j)},
\end{split}
\end{equation}

\begin{equation}
R_{(\alpha,i),b} 
= - P_T^{[1]}(\xi_i^2) e^{\alpha}_{(i)} \wedge e^{b},
\end{equation}
\begin{equation}
R_{(m_i+\alpha,i),b} 
= -P_T^{[1]}(\xi_i^2) e^{m_i+\alpha}_{(i)} \wedge
e^{b},
\end{equation}
\begin{equation}
R_{ab}
= \tilde{R}_{ab}
- P_T^{[1]}(0) \, e^{a} \wedge e^{b}.
\end{equation}


\section{Spin connections and curvature $2$-forms
for the special type}

In this appendix, we give explicit forms of the spin connections
and the Riemann curvature $2$-forms for the metric of the special
type: we consider the metric \eqref{gCKY} 
with a special metric $g^{(0)}_{\mathrm{special}}$
given by \eqref{g0sp}.

Let us introduce an orthonormal frame
$\{\hat{e}^{A} \}=\{\hat{e}^{\mu}, 
\hat{e}^{n+\mu}, \hat{e}^{2n+1}, 
\hat{e}^{\hat{\alpha}}_{(i)} \}$ :
\begin{equation}
\hat{e}^{\mu}= e^{\mu},
~~\hat{e}^{n+\mu}=e^{n+\mu},~~
\hat{e}^{2n+1}=\sqrt{S}\sum_{k=0}^{n} \sigma_k \theta_k,~~
\hat{e}^{\hat{\alpha}}_{(i)}= e^{\hat{\alpha}}_{(i)},
\end{equation}
where $S=c/\sigma_n$.
Then, the dual vector fields  are 
\begin{eqnarray}
\hat{e}_{\mu} &=& e_{\mu},~~~
\hat{e}_{n+\mu}= e_{n+\mu}
+ \frac{(-1)^n}{U_{\mu} \sqrt{P_{\mu}} x_{\mu}^2} 
\frac{\partial}{\partial \psi_n}, \\
\hat{e}_{2n+1}&=& 
\left( \frac{S}{c^2} \right)^{1/2} 
\frac{\partial}{\partial \psi_n},~~~
\hat{e}_{\hat{\alpha}}^{(i)}= \left( 
\prod_{\mu=1}^n (x_{\mu}^2-\xi_i^2) \right)^{-1/2} 
\bar{e}_{\hat{\alpha}}^{(i)},\nonumber
\end{eqnarray}
where
\begin{equation} \label{sbei}
\bar{e}_{\hat{\alpha}}^{(i)}
= \tilde{e}_{\hat{\alpha}}^{(i)}
+2 A^{(i)}_{\hat{\alpha}} 
\sum_{k=0}^{n}(-1)^{n+k} \xi_i^{2(n-k)-1} 
\frac{\partial}{\partial \psi_k}. 
\end{equation}
The connection 1-forms $\omega_{AB}$ are given by
\begin{eqnarray}
\hat{\omega}_{\mu \nu} &=& \omega_{\mu \nu},~~~
\hat{\omega}_{n+\mu, n+\nu}
=\omega_{n+\mu, n+\nu}, \nonumber\\
\hat{\omega}_{\mu, n+\nu}
&=& \omega_{\mu, n+\nu}+\delta_{\mu \nu} 
\frac{\sqrt{S}}{x_{\mu}} \hat{e}^{2 n+1},\\
\hat{\omega}_{\mu, 2n+1} 
&=& \frac{\sqrt{S}}{x_{\mu}} \hat{e}^{n+\mu}
-\frac{\sqrt{P_{\mu}}}{x_{\mu}} \hat{e}^{2n+1},~~~
\hat{\omega}_{n+\mu, 2n+1} 
= -\frac{\sqrt{S}}{x_{\mu}}\hat{e}^{\mu}, \nonumber
\end{eqnarray}
\begin{eqnarray}
\hat{\omega}_{\mu, (\alpha,i)}
&=& \omega_{\mu, (\alpha,i)} ,~~~
\hat{\omega}_{\mu, (m_i+\alpha,i)}
= \omega_{\mu, (m_i+\alpha,i)}, \\
\hat{\omega}_{n+\mu, (\alpha,i)}
&=&\omega_{n+\mu, (\alpha,i)} ,~~~
\hat{\omega}_{n+\mu, (m_i+\alpha,i)}
=\omega_{n+\mu, (m_i+\alpha,i)},\nonumber
\end{eqnarray}
\begin{equation}
\hat{\omega}_{2n+1, (\alpha,i)}
= - \frac{\sqrt{S}}{\xi_{i}}
e^{m_i+\alpha}_{(i)}, \qquad
\hat{\omega}_{2n+1, (m_i+\alpha,i)} 
=\frac{\sqrt{S}}{\xi_{i}}e^{\alpha}_{(i)},
\end{equation}
\begin{equation}
\begin{split}
\hat{\omega}_{(\alpha,i), ( \beta,j)} &= 
\omega_{(\alpha,i), ( \beta,j)} , \cr
\hat{\omega}_{(\alpha,i), (m_j+\beta,j)}
&= \omega_{(\alpha,i), (m_j+\beta,j)}
+\delta_{ij} \delta_{\alpha \beta} 
\frac{\sqrt{S}}{\xi_i} \hat{e}^{2n+1}, \cr
\hat{\omega}_{(m_i+\alpha, i), (m_j+\beta,j)} 
&=\omega_{(m_i+\alpha, i), (m_j+\beta,j)}.
\end{split}
\end{equation}
We use a function
\begin{equation}
\hat{P}_T^{[k]}(t):= \sum_{\mu=1}^n 
\frac{P_{\mu}}{(x_{\mu}^2 - t)^k}
+ \frac{S}{(-t)^k}
\end{equation}
with $\hat{P}_T= \hat{P}_T^{[0]}(t)$. 
Then, the curvature two forms 
$\hat{R}_{AB}$ are obtained by the replacements
$P_T \rightarrow \hat{P}_T$ and 
$P_T^{[k]} \rightarrow \hat{P}_T^{[k]}$ in the general type
together with
\begin{eqnarray}
\hat{R}_{\mu,2n+1}&=&-\frac{1}{2 x_{\mu}} 
\frac{\partial \hat{P}_T}{\partial x_{\mu}}
\hat{e}^{\mu} \wedge \hat{e}^{2n+1},~~~\hat{R}_{n+\mu,2n+1}=
-\frac{1}{2 x_{\mu}} \frac{\partial \hat{P}_T}{\partial x_{\mu}}
\hat{e}^{n+\mu} \wedge \hat{e}^{2n+1}, \\
\hat{R}_{(\alpha, i), 2n+1} &=& 
- \hat{P}_{T}^{[1]}(\xi_i^2) \hat{e}_{(i)}^{\alpha} \wedge \hat{e}^{2n+1},~~
\hat{R}_{(m_i+\alpha, i), 2n+1} = 
- \hat{P}_{T}^{[1]}(\xi_i^2) \hat{e}_{(i)}^{m_i+\alpha} 
\wedge \hat{e}^{2n+1}.
\end{eqnarray}

\section{$D=2n'+1$ Kerr-de Sitter black hole}

In this section of Appendix, 
we will show that the particular subclass of the
special case 
metric \eqref{spsp} indeed represent 
the odd-dimensional Kerr-de Sitter metric 
with partially equal angular momenta.
 
$D=2n'+1$ Kerr-de Sitter metric \cite{GLPP1,GLPP2} is given by
\beq \label{KdS}
g = d\bar{s}^2 + \frac{2M}{U}
\left( W dt + F dr - \sum_{I=1}^{n'} 
\frac{a_I \mu_I^2}{1 + \lambda a_I^2} d \phi_I \right)^2,
\eeq
where
\beq
\begin{split}
d\bar{s}^2 &= - W ( 1 - \lambda r^2) dt^2
+ F dr^2 + \sum_{I=1}^{n'} \frac{r^2 + a_I^2}{1 + \lambda a_I^2}
( d \mu_I^2 + \mu_I^2 d \phi_I^2) \cr
& + \frac{\lambda}{W(1 - \lambda r^2)}
\left( \sum_{I=1}^{n'} 
\frac{(r^2+a_I^2) \mu_I d\mu_I}{1 + \lambda a_I^2}
\right)^2,
\end{split}
\eeq
\beq
\sum_{I=1}^{n'} \mu_I^2 = 1, \qquad
U = \sum_{I=1}^{n'} \frac{\mu_I^2}{r^2 + a_I^2} \prod_{J=1}^{n'}
(r^2 + a_J^2).
\eeq
\beq
W = \sum_{I=1}^{n'} \frac{\mu_I^2}{1 + \lambda a_I^2}, \qquad
F = \frac{r^2}{1 - \lambda r^2}
\sum_{I=1}^{n'} \frac{\mu_I^2}{r^2 + a_I^2},
\eeq
The above Kerr-de Sitter metric has $n'$ angular momenta
$a_I$ $(I=1,2,\dotsc, n')$. In the following part, 
we require that the angular momenta are partially equal, namely
$(m_i+1)$ of them are chosen to $\xi_i$ $(i=1,2,\dotsc, n)$:
\beq
\begin{split}
a_i &= \xi_1, \qquad i=1,2,\dotsc, (m_1+1), \cr
a_{m_1+1+i} &= \xi_2, \qquad i=1,2,\dotsc, (m_2+1), \cr
\dotsm & \cr
a_{m_1 + m_2 + \dotsm + m_{n-1} + n-1 + i} &= \xi_n, \qquad
i=1,2,\dotsc, (m_n+1).
\end{split}
\eeq
Here $\xi_i \neq 0$ ($\forall i$) and 
$\xi_i \neq \xi_j$ ($\forall i \neq \forall j$).
Hence, 
\beq
n' = \sum_{i=1}^n ( m_i + 1) 
= n + |m|, \qquad |m|:= \sum_{i=1}^n m_i. 
\eeq
One will see that to each non-zero constant $\xi_i$,
there corresponds to the complex projective space
$\mathbb{CP}^{m_i}$. For this purpose, it is 
convenient to change the coordinates $\mu_I$
into $\{ r_i, u_{i,j} \}$:
\beq
\begin{split} \label{CT1}
\mu_i &= r_1 u_{1,i}, \qquad i=1,2,\dotsc, (m_1+1), \cr
\mu_{m_1+1+i} &= r_2 u_{2,i}, \qquad i=1,2,\dotsc, (m_2+1), \cr
\dotsm & \cr
\mu_{m_1+m_2+\dotsm + m_{n-1}+n-1 + i} &= r_{n} u_{n,i}, \qquad
i=1,2,\dotsc, (m_n+1),
\end{split}
\eeq
with constraints
\beq
\sum_{j=1}^{m_i+1} u_{i,j}^2 = 1, \qquad
i=1,2,\dotsc, n, \qquad
\sum_{i=1}^n r_i^2 = 1.
\eeq
Also the angular variables $\phi_I$ are renamed
$\varphi_{i,j}$:
\begin{align} \label{CT2}
\varphi_{1,i}&:= \phi_i,&  i&=1,2,\dotsc, (m_1+1), \cr
\varphi_{2,i}&:= \phi_{m_1+1+i},& i&=1,2,\dotsc, (m_2+1), \cr
\dotsm & \cr
\varphi_{n,i}&:= \phi_{m_1+m_2+\dotsm+m_{n-1} + n-1 + i},& 
i&=1,2,\dotsc, (m_n+1).
\end{align}
Then
\beq
\begin{split}
d\bar{s}^2 &= - W( 1 - \lambda r^2) dt^2
+ F dr^2 
+ \sum_{i=1}^n \left( \frac{r^2 + \xi_i^2}{1 + \lambda \xi_i^2}
\right) \Bigl( dr_i^2 + r_i^2 d\Omega_{i, (2m_i+1)}^2
\Bigr) \cr
& + \frac{\lambda}{W(1 - \lambda r^2)}
\left( \sum_{i=1}^n 
\frac{(r^2+\xi_i^2) r_i dr_i}{1 + \lambda \xi_i^2}
\right)^2,
\end{split}
\eeq
where $d\Omega_{i, (2m_i+1)}^2$ is the metric 
on the sphere $S^{2m_i+1}$ with unit radius:
\beq
d \Omega_{i, (2m_i+1)}^2 = \sum_{j=1}^{m_i+1} \bigl(
du_{i,j}^2 + u_{i,j}^2 d \varphi_{i,j}^2 \bigr), \qquad
\sum_{j=1}^{m_i+1} u_{i,j}^2 = 1.
\eeq
In a local coordinate patch where $u_{i,m_i+1}\neq 0$, 
we take
\beq \label{CT3}
\begin{split}
\varphi_{i,j} &=: \psi_i + \chi_{i,j}, \qquad j=1,2,\dotsc, m_i, \cr
\varphi_{i,m_i+1} &=: \psi_i.
\end{split}
\eeq
We get the Hopf fibration of $S^{2m_i+1}$,
a $U(1)$  bundle over $\mathbb{CP}^{m_i}$:
\beq
d \Omega_{(2m_i+1)}^2 = ( d \psi_i - 2 A_i)^2 + d \Sigma_{i,(m_i)}^2,
\eeq
where $d \Sigma_{i,(m_i)}^2$ is a real form of 
the Fubini-Study metric on $\mathbb{CP}^{m_i}$
\beq
d\Sigma_{i,(m_i)}^2 = \sum_{j=1}^{m_i+1} d u_{i,j}^2
+ \sum_{j=1}^{m_i} u_{i,j}^2 d \chi_{i,j}^2 
- \sum_{j=1}^{m_i} \sum_{k=1}^{m_i}
u_{i,j} u_{i,k} d \chi_{i,j} d \chi_{i,k},
\eeq
and $A_i$ is a potential for 
the corresponding K\"{a}hler form $J^{(i)} = d A_i$,
\beq
A_i = - \frac{1}{2} \sum_{j=1}^{m_i} u_{i,j}^2 d \chi_{i,j}.
\eeq
The Fubini-Study metric $d \Sigma_{i,(m_i)}^2$ has 
the cosmological constant $2(m_i+1)$.

Now the metric \eqref{KdS} with partially equal momenta
is written as the following form:
\beq \label{ds2odd}
\begin{split}
g &= \sum_{i=1}^n \left( \frac{r^2 + \xi_i^2}{1 + \lambda \xi_i^2}
\right)^2 d r_i^2 + \frac{\lambda}{W(1 - \lambda r^2)}
\left( \sum_{i=1}^n \frac{(r^2+\xi_i^2) r_i dr_i}{1 + \lambda \xi_i^2}
\right)^2 \cr
& + F dr^2 - W(1 - \lambda r^2) dt^2 + \sum_{i=1}^{n}
\left( \frac{r^2 + \xi_i^2}{1 + \lambda \xi_i^2} \right)
r_i^2 ( d \psi_i - 2 A_i )^2 \cr
& + \frac{2M}{U} \left( W dt - \sum_{i=1}^n
\left( \frac{\xi_i r_i^2}{1 + \lambda \xi_i^2} \right)
( d\psi_i - 2 A_i) + F dr \right)^2 \cr
& + \sum_{i=1}^n \left( \frac{r^2 + \xi_i^2}{1 + \lambda \xi_i^2}
\right) r_i^2 d \Sigma_{i,(m_i)}^2,
\end{split}
\eeq
with a constraint
\beq \label{constri}
\sum_{i=1}^n r_i^2 = 1.
\eeq
The functions in the metric are given by
\beq
W = \sum_{i=1}^n \frac{r_i^2}{1 + \lambda \xi_i^2}, \qquad
F = \frac{r^2}{1 - \lambda r^2} \sum_{i=1}^n 
\frac{r_i^2}{r^2 + \xi_i^2},
\qquad
U = \sum_{i=1}^n \frac{r_i^2}{r^2 + \xi_i^2} 
\prod_{j=1}^n (r^2+\xi_j^2)^{m_j+1}.
\eeq

The constraint \eqref{constri} can be solved by setting
\beq \label{CT4}
r_i^2 = \prod_{\mu=1}^{n-1} ( \xi_i^2 - x_{\mu}^2)
\prod_
{\stackrel{\scriptstyle j=1}{(j \neq i)}}^n 
( \xi_i^2 - \xi_j^2)^{-1}, 
\qquad
i=1,2,\dotsc, n.
\eeq
Then the above functions turn into the following forms:
\beq
W = \prod_{\mu=1}^{n-1} ( 1 + \lambda x_{\mu}^2)
\prod_{i=1}^{n} ( 1 + \lambda \xi_i^2)^{-1}, \qquad
F = r^2 ( 1 - \lambda r^2)^{-1} 
\prod_{\mu=1}^{n-1} ( r^2 + x_{\mu}^2)
\prod_{i=1}^{n} ( r^2 + \xi_i^2)^{-1},
\eeq
\beq
U = \prod_{\mu=1}^{n-1} (r^2 + x_{\mu}^2)
\prod_{i=1}^n (r^2 + \xi_i^2)^{m_i}.
\eeq
One can check that
\beq
 \sum_{i=1}^n \left( 
\frac{r^2 + \xi_i^2}{1 + \lambda \xi_i^2}
\right)^2 d r_i^2 
+ \frac{\lambda}{W(1 - \lambda r^2)}
\left( \sum_{i=1}^n 
\frac{(r^2+\xi_i^2) r_i dr_i}{1 + \lambda \xi_i^2}
\right)^2  = \sum_{\mu=1}^{n-1} 
\frac{dx_{\mu}^2}{P_{\mu}},
\eeq
where (for $\mu=1,2,\dotsc, n-1$)
\beq
P_{\mu} := - \frac{( 1 + \lambda x_{\mu}^2)}
{x_{\mu}^{2} U_{\mu}} 
\prod_{i=1}^n (x_{\mu}^2 - \xi_i^2), \qquad
U_{\mu} = ( x_{\mu}^2 + r^2) 
\prod_{\stackrel{\scriptstyle \nu =1}{(\nu \neq \mu)}}^{n-1}
(x_{\mu}^2 - x_{\nu}^2).
\eeq
It is convenient to introduce coordinates $(\tilde{t}, \tilde{\psi}_i,
\tilde{r})$ by
\beq \label{CT5}
\begin{split}
d\tilde{t}&:= dt - \frac{2Mr^2}{(1 - \lambda r^2) V(r)} dr, \cr
d\tilde{\psi}_i &= d \psi_i 
- \frac{2M \xi_i r^2}{(r^2+\xi_i^2) V(r)} dr, \cr
d\tilde{r} &:= dr,
\end{split}
\eeq
where
\beq
V(r) = - 2 M r^2 + (1 - \lambda r^2) \prod_{i=1}^n 
( r^2 + \xi_i^2)^{m_i+1}.
\eeq 
With some work, we can prove that the following relation holds:
\beq
\begin{split}
&  F dr^2 - W(1 - \lambda r^2) dt^2 + \sum_{i=1}^{n}
\left( \frac{r^2 + \xi_i^2}{1 + \lambda \xi_i^2} \right)
r_i^2 ( d \psi_i - 2 A_i )^2 \cr
& + \frac{2M}{U} \left( W dt - \sum_{i=1}^n
\left( \frac{\xi_i r_i^2}{1 + \lambda \xi_i^2} \right)
( d\psi_i - 2 A_i) + F dr \right)^2 \cr
&= - \frac{d\tilde{r}^2}{P_n}
- W(1 - \lambda \tilde{r}^2) d \tilde{t}\, {}^2
+ \sum_{i=1}^{n}
\left( \frac{\tilde{r}^2 + \xi_i^2}{1 + \lambda \xi_i^2} \right)
r_i^2 ( d \tilde{\psi}_i - 2 A_i )^2 \cr
& + \frac{2M}{U} \left( W d\tilde{t} - \sum_{i=1}^n
\left( \frac{\xi_i r_i^2}{1 + \lambda \xi_i^2} \right)
( d\tilde{\psi}_i - 2 A_i) \right)^2,
\end{split}
\eeq
where
\beq
P_n = \frac{\tilde{X}_n(\tilde{r})}{U_n},
\eeq
\beq
\tilde{X}_n(\tilde{r}) = (-1)^n V(\tilde{r}) 
\tilde{r}^{-2} \prod_{i=1}^n (\tilde{r}^2 + \xi_i^2)^{-m_i}, \qquad
U_n=(-1)^{n-1}\prod_{\mu=1}^{n-1}(\tilde{r}^2 + x_{\mu}^2 ).
\eeq
We rescale the Fubini-Study metric such that
\beq
\left( \frac{r^2 + \xi_i^2}{1 + \lambda \xi_i^2}
\right) r_i^2 d \Sigma_{i,(m_i)}^2
= \prod_{\mu=1}^n ( x_{\mu}^2 - \xi_i^2) g^{(i)}, \qquad
i=1,2,\dotsc, n,
\eeq
with $\tilde{r} = i x_n$, namely
\beq
g^{(i)}:= (-1)^n ( 1 + \lambda \xi_i^2)^{-1}
\prod_{\stackrel{\scriptstyle j=1}{(j \neq i)}}^n
(\xi_i^2 - \xi_j^2)^{-1} d \Sigma_{i,(m_i)}^2.
\eeq 
The Fubini-Study metric $g^{(i)}$ on $\mathbb{CP}^{m_i}$
has the cosmological constant
\beq
\lambda^{(i)} = (-1)^n 2 (m_i+1) (1 + \lambda \xi_i^2)
\prod_{\stackrel{\scriptstyle j=1}{(j \neq i)}}^n
(\xi_i^2 - \xi_j^2).
\eeq
We also rescale the corresponding potential $A_i$ 
and the angle $\tilde{\psi}_i$ as follows:
\beq
\tilde{\psi}_i':=(-1)^n ( 1 + \lambda \xi_i^2)^{-1}
\prod_{\stackrel{\scriptstyle j=1}{(j \neq i)}}^n
(\xi_i^2 - \xi_j^2)^{-1} \tilde{\psi}_i,
\eeq
\beq
A_i':= (-1)^n ( 1 + \lambda \xi_i^2)^{-1}
\prod_{\stackrel{\scriptstyle j=1}{(j \neq i)}}^n
(\xi_i^2 - \xi_j^2)^{-1} A_i.
\eeq
For simplicity, let
\beq
\xi_0^2:= - \frac{1}{\lambda}, \qquad
d \tilde{t} =: - \frac{1}{\xi_0} \prod_{i=1}^n
( \xi_i^2 - \xi_0^2) d \tilde{\psi}_0',
\eeq
\beq
\vartheta_0:= d \tilde{\psi}_0', \qquad
\vartheta_i:= d \tilde{\psi}_i' - 2 A_i', \qquad
i=1,2,\dotsc, n.
\eeq
We can check that
\beq
\begin{split} \label{mS}
& - W(1 - \lambda \tilde{r}^2) d \tilde{t}\, {}^2
+ \sum_{i=1}^{n}
\left( \frac{\tilde{r}^2 + \xi_i^2}{1 + \lambda \xi_i^2} \right)
r_i^2 ( d \tilde{\psi}_i - 2 A_i )^2 \cr
& + \frac{2M}{U} \left( W d\tilde{t} - \sum_{i=1}^n
\left( \frac{\xi_i r_i^2}{1 + \lambda \xi_i^2} \right)
( d\tilde{\psi}_i - 2 A_i) \right)^2 \cr
&= \sum_{\mu=1}^n P_{\mu}
\left( \sum_{\hat{i}=0}^{n} \xi_{\hat{i}}
\prod_{\stackrel{\scriptstyle \nu=1}{(\nu \neq \mu)}}^n
( x_{\nu}^2 - \xi_{\hat{i}}^2) \vartheta_{\hat{i}} \right)^2 
+ \frac{c}{\sigma_n}
\left( \sum_{\hat{i}=0}^n
\frac{1}{\xi_{\hat{i}}}
\prod_{\nu=1}^n ( x_{\nu}^2 - \xi_{\hat{i}} ) \vartheta_{\hat{i}}
\right)^2,
\end{split}
\eeq
where
\beq
c = - \prod_{i=1}^n \xi_i^2.
\eeq
Let us define $1$-forms $\theta_k$ by
\beq
\theta_k:= \sum_{\hat{i}=0}^n (-1)^{n-k} \xi_{\hat{i}}^{2(n-k)-1}
\vartheta_{\hat{i}}, \qquad
k=0,1,\dotsc, n.
\eeq
They satisfy the equations \eqref{eqth} for $N=n$ and $\varepsilon=1$:
\beq
d \theta_k  + 2 \sum_{i=1}^n (-1)^{n-k} \xi_{i}^{2(n-k)-1}
\omega^{(i)} = 0, \qquad
\omega^{(i)} := d A_i', \qquad
(k=0,1,\dotsc, n).
\eeq
Note that 
\beq \label{thetak}
\begin{split}
\theta_k &= \sum_{\hat{i}=0}^n (-1)^{n-k} \xi_{\hat{i}}^{2(n-k)-1}
\vartheta_{\hat{i}} \cr
&= - \frac{\lambda^k}{\displaystyle \prod_{i=1}^n ( 1 + \lambda \xi_i^2)}
d \tilde{t} + (-1)^{n-k} \sum_{i=1}^n \xi_i^{2(n-k)-1}
( d \tilde{\psi}_i' - 2 A_i') \cr
& = - \frac{\lambda^k}{\displaystyle \prod_{i=1}^n ( 1 + \lambda \xi_i^2)}
d \tilde{t}
+ (-1)^k \sum_{i=1}^n \frac{\xi_i^{2(n-k)-1}}{\displaystyle
(1 + \lambda \xi_i^2) \prod_{\stackrel{\scriptstyle j=1}{(j \neq i)}}^n
( \xi_i^2 - \xi_j^2)}
( d \tilde{\psi}_i - 2 A_i ).
\end{split}
\eeq
Using these $1$-forms $\theta_k$,
summations within brackets in the
last line of \eqref{mS} can be rewritten as follows:
\beq
\sum_{\hat{i}=0}^{n} \xi_{\hat{i}}
\prod_{\stackrel{\scriptstyle \nu=1}{(\nu \neq \mu)}}^n
( x_{\nu}^2 - \xi_{\hat{i}}^2) \vartheta_{\hat{i}} 
= - \sum_{k=0}^{n-1} \sigma_k(\hat{x}_{\mu}) \theta_k,
\eeq
\beq
\sum_{\hat{i}=0}^n
\frac{1}{\xi_{\hat{i}}}
\prod_{\nu=1}^n ( x_{\nu}^2 - \xi_{\hat{i}} ) \vartheta_{\hat{i}}
= \sum_{k=0}^n \sigma_k \theta_k.
\eeq
Combining these relations,
the metric \eqref{ds2odd} finally becomes the metric \eqref{spsp}: 
\beq
g = \sum_{\mu=1}^n \frac{dx_{\mu}^2}{P_{\mu}}
+ \sum_{\mu=1}^n P_{\mu}
\left[ \sum_{k=0}^{n-1} \sigma_k(\hat{x}_{\mu}) \theta_k \right]^2
+ \frac{c}{\sigma_n}
\left[ \sum_{k=0}^n \sigma_k \theta_k \right]^2
+ \sum_{i=1}^n \prod_{\mu=1}^n ( x_{\mu}^2 - \xi_i^2) g^{(i)},
\eeq
where
\beq
P_{\mu} = \frac{X_{\mu}(x_{\mu})}{\displaystyle x_{\mu}
\prod_{i=1}^n ( x_{\mu}^2 - \xi_i^2)^{m_i} U_{\mu}},
\qquad
U_{\mu} = \prod_{\stackrel{\scriptstyle \nu=1}{(\nu \neq \mu)}}^n
( x_{\mu}^2 - x_{\nu}^2),
\qquad
c = - \prod_{i=1}^n \xi_i^2,
\eeq
\beq
X_{\mu}(x_{\mu}) = x_{\mu}
\left( (-1)^{n+|m|-1} 2M \delta_{\mu,n} - 
\left( 1 + \lambda x_{\mu}^2 \right) x_{\mu}^{-2} 
\prod_{i=1}^n ( x_{\mu}^2 - \xi_i^2)^{m_i+1}
\right).
\eeq

Hence, we have seen that by the coordinate transformations
\eqref{CT1}, \eqref{CT2}, \eqref{CT3}, \eqref{CT4} and
\eqref{CT5}, the odd-dimensional
general Kerr-de Sitter metric \eqref{KdS}
with equal angular momenta turns into the subfamily
of the special case of the generalized Kerr-NUT-de Sitter
metric \eqref{spsp}. Here all angular momenta take non-zero values.

Let us summarize the special type metric \eqref{spsp}
with $N=n$, $m^{(0)}=0$ and $\varepsilon=1$:
the base space is the direct product of
complex projective spaces
\beq
B= M^{(1)} \times M^{(2)} \times
\dotsm \times M^{(n)} = 
\mathbb{CP}^{m_1} \times \mathbb{CP}^{m_2}
\times \dotsm \times \mathbb{CP}^{m_n}
\eeq
and the fiber over $B$ is a $(2n+1)$-dimensional 
Kerr-de Sitter space.

\subsection{From special type to odd dimensional general type}

By sending one of constant eigenvalues, say $\xi_n$, to zero,
the metric of special type \eqref{spsp} goes to a metric of
general type with
$N = n-1$, $m^{(0)} = 2m_n+1$ and $\varepsilon=0$.
The base space is the direct product of
complex projective spaces and a $(2m_n+1)$-dimensional
sphere $S^{2m_n+1}$
\beq
B = M^{(1)} \times M^{(2)} \times \dotsm
\times M^{(n-1)} \times M^{(0)}
= \mathbb{CP}^{m_1} \times \mathbb{CP}^{m_2}
\times \dotsm \times \mathbb{CP}^{m_{n-1}} \times
S^{2m_n+1},
\eeq
and the fiber over $B$ is a $2n$-dimensional Kerr-de Sitter space.

For $k=0,1,\dotsc, n-1$, 
the $1$-form $\theta_k$ \eqref{thetak}
has a smooth $\xi_n \rightarrow 0$
limit:
\beq
\begin{split}
\tilde{\theta_k}&:= \lim_{\xi_n \rightarrow 0} \theta_k \cr 
&= - \lambda^k 
\prod_{i=1}^{n-1} ( 1 + \lambda \xi_j^2)^{-1} d \tilde{t} \cr
& + (-1)^k \sum_{i=1}^{n-1} \xi_i^{2(n-1-k) -1}
( 1 + \lambda \xi_i^2)^{-1}
\prod_{\stackrel{\scriptstyle j=1}{(j \neq i)}}^{n-1}
( \xi_i^2 - \xi_j^2)^{-1} ( d \tilde{\psi}_i - 2 A_i ),
\end{split}
\eeq
while $\theta_n$ has a singular limit
\beq
\theta_n = - \frac{1}{\xi_n} \prod_{j=1}^{n-1} \xi_j^{-2}
( d \psi_n - 2 A_n) + O(1).
\eeq
Here we have used
\beq
d \tilde{\psi}_n = d \psi_n + O(\xi_n).
\eeq
The leading term of $\sum_{k=0}^n \sigma_k \theta_k$
is $\sigma_n \theta_n$. Hence it follows that
\beq
\lim_{\xi_n \rightarrow 0} \frac{c}{\sigma_n}
\left[ \sum_{k=0}^n \sigma_k \theta_k \right]^2
= - \sigma_n \left( \prod_{j=1}^{n-1} \xi_j^{-2} \right)
( d \psi_n - 2 A_n)^2.
\eeq
Also
\beq
\lim_{\xi_n \rightarrow 0} g^{(n)}
= - \left( \prod_{j=1}^{n-1} \xi_j^{-2} \right) d
\Sigma_{n,(m_n)}^2.
\eeq
In the $\xi_n \rightarrow 0$ limit, the metric \eqref{spsp}
becomes
\beq \label{sptogen}
g = \sum_{\mu=1}^n \frac{dx_{\mu}^2}{P_{\mu}}
+ \sum_{\mu=1}^n P_{\mu}
\left[ \sum_{k=0}^{n-1} \sigma_k(\hat{x}_{\mu}) 
\tilde{\theta}_k \right]^2
+ \sum_{i=1}^{n-1} \prod_{\mu=1}^{n} ( x_{\mu}^2 - \xi_i^2)
g^{(i)} - \sigma_n
\left( \prod_{j=1}^{n-1} \xi_j^{-2} \right) 
d \Omega_{n,(2m_n+1)}^2,
\eeq
where
\beq
d \Omega_{n,(2m_n+1)}^2 = ( d \psi_n - 2 A_n)^2
+ d \Sigma_{n,(m_n)}^2,
\eeq
\beq
P_{\mu} = \frac{X_{\mu}(x_{\mu})}
{\displaystyle (x_{\mu})^{2m_n+1}
\prod_{i=1}^{n-1} ( x_{\mu}^2 - \xi_i^2)^{m_i} U_{\mu} },
\eeq
\beq
X_{\mu}(x_{\mu}) = x_{\mu}\left(
(-1)^{(1/2)(D-1)-1} 2M \delta_{\mu, n} 
- \left( 1 + \lambda x_{\mu}^2 \right)
x_{\mu}{}^{2m_n}
\prod_{i=1}^{n-1} (x_{\mu}^2 - \xi_i^2)^{m_i+1}
\right).
\eeq
Therefore, we get the metric \eqref{evng} for odd $m^{(0)}$.
Hence we have shown that the odd dimensional case of
\eqref{evng} represents the odd dimensional 
Kerr-de Sitter black hole
with partially equal angular momenta and with some
zero angular momenta.

Remark. The function $\chi(x)= \sum_{i=-1}^{n} \alpha_i x^{2i}$
\eqref{chisp} 
has a smooth limit into $\chi(x) = \sum_{i=0}^n \alpha_i x^{2i}$:
$\lim_{\xi_n \rightarrow 0} \alpha_{-1} = 0$, and
\beq
\begin{split}
\chi(x) = \sum_{i=0}^{n} \alpha_i x^{2i} 
&= -2 \sum_{i=1}^{n-1} (m_i+ 1) ( 1 + \lambda \xi_i^2) x^2
\prod_{\stackrel{\scriptstyle j=1}{(j \neq i)}}^{n-1}
( x^2 - \xi_j^2) \cr
& - 2 \Bigl( m_n ( 1 + \lambda x^2) + ( n + |m|) \lambda x^2
\Bigr) \prod_{i=1}^{n-1} ( x^2 - \xi_i^2),
\end{split}
\eeq
where
\beq
\alpha_0 = (-1)^{n} 2 m_n \prod_{j=1}^{n-1} \xi_j^2, \qquad
| m | = \sum_{i=1}^{n-1} m_i.
\eeq

\section{$D=2n'$ Kerr-de Sitter black hole}

The general Kerr-de Sitter metric in $D=2n'$ can be obtained from 
that of $D=2n'+1$ by setting \cite{GLPP1,GLPP2}
\beq \label{oddtoeven}
\phi_{n'}=0, \qquad
a_{n'}=0, \qquad M \rightarrow M r.
\eeq
The metric is given by
\beq \label{evn}
g = d\bar{s}^2 + \frac{2Mr}{U}
\left( W dt + F dr - \sum_{I=1}^{n'-1}
 \frac{a_I \mu_I^2}{1 + \lambda a_I^2} d \phi_I \right)^2,
\eeq
\beq
\begin{split}
d\bar{s}^2 &= - W ( 1 - \lambda r^2) dt^2 + F dr^2
+ \sum_{I=1}^{n'} \frac{r^2 + a_I^2}{1 + \lambda a_I^2}
d \mu_I^2 + \sum_{I=1}^{n'-1} \frac{r^2 + a_I^2}{1 + \lambda a_I^2}
\mu_I^2 d\phi_I^2 \cr
& + \frac{\lambda}{W(1 - \lambda r^2)}
\left( \sum_{I=1}^{n'} \frac{(r^2+a_I^2) \mu_I d \mu_I}{1 + \lambda a_I^2}
\right)^2,
\end{split}
\eeq
\beq
W = \sum_{I=1}^{n'} \frac{\mu_I^2}{1 + \lambda a_I^2}, \qquad
F = \frac{r^2}{1 - \lambda r^2} \sum_{I=1}^{n'} 
\frac{\mu_I^2}{r^2 + a_I^2},
\eeq
\beq
U = r^2 \left( \sum_{I=1}^{n'} \frac{\mu_I^2}{r^2 + a_I^2} \right)
\prod_{J=1}^{n'-1} (r^2 + a_J^2), \qquad
\sum_{I=1}^{n'} \mu_I^2 = 1.
\eeq  
Remark. For $D=2n'$, $U/r$ here is written as $U$ in \cite{GLPP1,GLPP2}. 

Then taking 
\beq
n' = n + |m| = n + \sum_{i=1}^{n} m_i, \qquad
m_n := 0,
\qquad
\xi_n = 0,
\eeq
with \eqref{oddtoeven}, the metric \eqref{evn} 
with partially equal angular momenta can be written as
\beq
\begin{split}
g &= \sum_{i=1}^n \left( 
\frac{r^2 + \xi_i^2}{ 1 + \lambda \xi_i^2} \right)
d r_i^2 + \frac{\lambda}{W(1 - \lambda r^2)}
\left( \sum_{i=1}^n \frac{(r^2 + \xi_i^2) r_i dr_i}{1 + \lambda \xi_i^2}
\right)^2 \cr
& + F dr^2 - W ( 1 - \lambda r^2) dt^2 
+ \sum_{i=1}^{n-1} \left( \frac{r^2+\xi_i^2}{1 + \lambda \xi_i^2}
\right) r_i^2 ( d\psi_i - 2 A_i)^2 \cr
& + \frac{2Mr}{U} \left( W dt
- \sum_{i=1}^{n-1} \left( \frac{\xi_i r_i^2}{1 + \lambda \xi_i^2}
\right)( d \psi_i - 2 A_i ) + F dr \right)^2 \cr
& + \sum_{i=1}^{n-1} \left( \frac{r^2 + \xi_i^2}{1 + \lambda \xi_i^2}
\right) r_i^2 d \Sigma_{i,(m_i)}^2,
\end{split}
\eeq
with a constraint
\beq
\sum_{i=1}^n r_i^2 = 1.
\eeq
This can be rewritten as the form
\beq \label{ggKdS}
g = \sum_{\mu=1}^n \frac{dx_{\mu}^2}{P_{\mu}}
+ \sum_{\mu=1}^n P_{\mu}
\left( \sum_{k=0}^{n-1} \sigma_k( \hat{x}_{\mu}) \theta_k \right)^2
+ \sum_{i=1}^{n-1} \prod_{\mu=1}^{n} ( x_{\mu}^2 - \xi_i^2) g^{(i)},
\eeq
by the coordinate transformations:
\beq
\begin{split}
d \tilde{t} &= dt - \frac{2Mr^3}{(1 - \lambda r^2) V(r)} dr, \cr
d \tilde{\psi}_i &= d\psi_i 
- \frac{2M \xi_i r^3}{(r^2+\xi_i^2) V(r)} dr, \cr
d \tilde{r} &= dr,
\end{split}
\eeq
\beq
V(r) = - 2 M r^3 
+ (1 - \lambda r^2) r^2 \prod_{i=1}^{n-1} ( r^2 + \xi_i^2)^{m_i+1},
\eeq 
\beq \label{evri}
r_i^2 = \prod_{\mu=1}^{n-1} ( \xi_i^2 - x_{\mu}^2)
\prod_{\stackrel{\scriptstyle j=1}{( j \neq i)}}^n
(\xi_i^2 - \xi_j^2)^{-1}, \qquad
i=1,2,\dotsc, n, \qquad
\tilde{r} = i x_n,
\eeq
\beq
g^{(i)} = (-1)^n ( 1 + \lambda \xi_i^2)^{-1} \xi_i^{-2}
\prod_{\stackrel{\scriptstyle j=1}{(j \neq i)}}^{n-1}
( \xi_i^2 - \xi_j^2)^{-1} d \Sigma_{i,(m_i)}^2, \qquad
i=1,2,\dotsc, n-1,
\eeq
\beq
\xi_0^2:= - \frac{1}{\lambda},
\eeq
\beq \label{thetak2}
\begin{split}
\theta_k &= \sum_{\hat{i}=0}^{n-1}
(-1)^{n-k} \xi_{\hat{i}}^{2(n-k)-1} \vartheta_{\hat{i}} \cr
&= - \lambda^k \prod_{j=1}^{n-1} ( 1 + \lambda \xi_j^2)^{-1} d
\tilde{t}
+ (-1)^{n-k} \sum_{i=1}^{n-1} \xi_i^{2(n-k)-1} ( d \tilde{\psi}'_i
- 2 A_i') \cr
&= - \lambda^k 
\prod_{j=1}^{n-1} ( 1 + \lambda \xi_j^2)^{-1} d
\tilde{t}
+ (-1)^k \sum_{i=1}^{n-1}
\xi_i^{2(n-1-k)-1} ( 1 + \lambda \xi_i^2)^{-1}
\prod_{\stackrel{\scriptstyle j=1}{(j \neq i)}}^{n-1}
( \xi_i^2 - \xi_j^2)^{-1} ( d \tilde{\psi}_i - 2 A_i).
\end{split}
\eeq
Here (for $\mu=1,2,\dotsc, n$)
\beq
P_{\mu} = \frac{X_{\mu}(x_{\mu})}
{\displaystyle \prod_{i=1}^{n-1} ( x_{\mu}^2 - \xi_i^2)^{m_i}
U_{\mu}}, \qquad
U_{\mu} = \prod_{\stackrel{\scriptstyle \nu=1}{(\nu \neq \mu)}}^n
( x_{\mu}^2 - x_{\nu}^2),
\eeq
\beq
X_{\mu}(x_{\mu})
= (-1)^{n+|m|-1} 2 M i x_{\mu} \delta_{\mu,n}
- ( 1 + \lambda x_{\mu}^2) 
\prod_{j=1}^{n-1}( x_{\mu}^2 - \xi_j^2)^{m_j+1}.
\eeq
This metric \eqref{ggKdS}
represents the general type with $m^{(0)}=0$, 
$\varepsilon=0$,
\beq
D = 2n + 2|m| = 2n + 2 \sum_{i=1}^{n-1} m_i,
\eeq
the base space is $(N=n-1)$
\beq \label{B1}
B = M^{(1)} \times M^{(2)} \times \dotsm \times M^{(n-1)}
= \mathbb{CP}^{m_1} \times \mathbb{CP}^{m_2}
\times \dotsm \times \mathbb{CP}^{m_{n-1}},
\eeq
and the fiber is a $2n$-dimensional Kerr-de Sitter spacetime.

The Fubini-Study metric $g^{(i)}$ $(i=1,2,\dotsc, n-1)$ 
has the cosmological constant
\beq
\lambda^{(i)} = (-1)^n 2(m_i+1) ( 1 + \lambda \xi_i^2) \xi_i^2 
\prod_{\stackrel{\scriptstyle j=1}{( j \neq i)}}^{n-1} ( \xi_i^2 - \xi_j^2).
\eeq

\subsection{Introducing $M^{(0)}$ by taking zero eigenvalue limit}

The base space \eqref{B1} has no Einstein subspace $M^{(0)}$
which corresponds to the zero eigenvalues of the CKY tensor.
In this subsection, we introduce $M^{(0)}$
by sending one of non-zero constant eigenvalues $\xi_i$ to zero.

Let us consider a limit such that one of the constant eigenvalues,
say
$\xi_{n-1}$, goes to zero.
Since $\xi_n=0$, naive $\xi_{n-1}\rightarrow 0$ limit make the coordinate
transformation \eqref{evri} singular.
It is convenient to rescale one of $x_{\mu}$. 
For definiteness, we set $x_{n-1} = \xi_{n-1} \rho$
and take $\xi_{n-1} \rightarrow 0$ limit keeping $\rho$ finite.
The coordinate transformation \eqref{evri} becomes
\beq
\begin{split}
r_i^2 &= \xi_i^{-2} \prod_{\mu=1}^{n-1} ( \xi^2 - x_{\mu}^2)
\left( \prod_{\stackrel{\scriptstyle j=1}{(j \neq i)}}^{n-1}
( \xi_i^2 - \xi_j^2)^{-1} \right), \qquad
i=1,2,\dotsc, n-2, \cr
r_{n-1}^2& = \prod_{\mu=1}^{n-2} x_{\mu}^2 
\left( \prod_{j=1}^{n-2} \xi_j^{-2} \right) ( 1 - \rho^2), \cr
r_{n}^2 &= \prod_{\mu=1}^{n-2} x_{\mu}^2 
\left( \prod_{j=1}^{n-2} \xi_j^{-2} \right) \rho^2.
\end{split}
\eeq
The range of $\rho$ is restricted to the interval 
$-1 \leq \rho \leq 1$.
We can see that
\beq
\lim_{\xi_{n-1} \rightarrow 0}
\frac{P_{n-1}}{\xi_{n-1}^2}
= - \frac{1}{\tilde{\sigma}_{n-1}} \prod_{i=1}^{n-2} \xi_i^2
( 1 - \rho^2), \qquad
\tilde{\sigma}_{n-1} := \left( \prod_{\mu=1}^{n-2} x_{\mu}^2 \right) x_n^2.
\eeq
\beq
\lim_{\xi_{n-1} \rightarrow 0}
P_{n-1} \left( \sum_{k=0}^{n-1} \sigma_k( \hat{x}_{n-1}) \theta_k
\right)^2 = - \tilde{\sigma}_{n-1}
\left( \prod_{j=1}^{n-2} \xi_j^{-2} \right)
( 1 - \rho^2) ( d \psi_{n-1} - 2 A_{n-1} )^2,
\eeq
\beq
\lim_{\xi_{n-1} \rightarrow 0}
\left( \frac{r^2 + \xi_{n-1}^2}{1 + \lambda \xi_{n-1}^2}
\right) r_{n-1}^2 d \Sigma_{n-1,(m_{n-1})}^2
= - \tilde{\sigma}_{n-1}
\left( \prod_{j=1}^{n-2} \xi_j^{-2} \right)
( 1 - \rho^2) d \Sigma_{n-1,(m_{n-1})}^2.
\eeq
Combining these relations, we have
\beq
\begin{split}
& \frac{dx_{n-1}^2}{P_{n-1}}
+ P_{n-1} \left( \sum_{k=0}^{n-1} \sigma_k( \hat{x}_{n-1})
\theta_k \right)^2 
+ \left( \frac{r^2 + \xi_{n-1}^2}{1 + \lambda \xi_{n-1}^2}
\right) r_{n-1}^2 d \Sigma_{n-1,(m_{n-1})}^2 \cr
& \rightarrow
- \tilde{\sigma}_{n-1}
\left( \prod_{j=1}^{n-2} \xi_j^{-2} \right) 
d \tilde{\Omega}_{n-1,(2m_{n-1}+2)}^2,
\end{split}
\eeq
where 
\beq
d \tilde{\Omega}_{n-1,(2m_{n-1}+2)}^2
:= \frac{d \rho^2}{1 - \rho^2}
+ (1 - \rho^2) d \Omega_{n-1,(2m_{n-1}+1)}^2 
\eeq
is the metric on the sphere $S^{2m_{n-1}+2}$ with 
unit radius.

For $\mu=1,2,\dotsc, n-2,n$, let
\beq
\prod_{\stackrel{\scriptstyle \nu=1}{(\nu \neq \mu, n-1)}}^n
( t - x_{\nu}^2) = 
\sum_{k=0}^{n-2} (-1)^k t^{n-2-k} 
\sigma_k(\hat{x}_{\mu}, \hat{x}_{n-1})=:
\sum_{k=0}^{n-2} (-1)^k t^{n-2-k} 
\tilde{\sigma}_k(\hat{x}_{\mu}).
\eeq
It holds that for $\mu \neq n-1$,
\beq
\sigma_k(\hat{x}_{\mu}) = \tilde{\sigma}_k(\hat{x}_{\mu})
+ x_{n-1}^2 \tilde{\sigma}_{k-1}(\hat{x}_{\mu}).
\eeq
Especially, for $k =n-1$, we have
\beq
\sigma_{n-1}(\hat{x}_{\mu}) 
= x_{n-1}^2 \tilde{\sigma}_{n-2} (\hat{x}_{\mu}) 
= \xi_{n-1}^2 \rho^2 \tilde{\sigma}_{n-2}(\hat{x}_{\mu})
= O(\xi_{n-1}^2).
\eeq
From the last expression of $\theta_k$ in \eqref{thetak2},
we can see that $\theta_k = O(1)$ for $k=0,1,\dotsc, n-2$
and $\theta_{n-1} = O(\xi_{n-1}^{-1})$. We have
\beq
\lim_{\xi_{n-1} \rightarrow 0} \sigma_{n-1}(\hat{x}_{\mu}) \theta_{n-1}
= 0, \qquad
\mu \neq n-1,
\eeq
and
\beq
\lim_{\xi_{n-1} \rightarrow 0}
\sum_{k=0}^{n-1} \sigma_k(\hat{x}_{\mu}) \theta_k
= \sum_{k=0}^{n-2} \tilde{\sigma}_k(\hat{x}_{\mu})
\tilde{\theta}_k, \qquad \mu \neq n-1,
\eeq
where
\beq
\tilde{\theta}_k
= - \lambda^k \prod_{j=1}^{n-2} (1 + \lambda \xi_j^2 )^{-1}
d \tilde{t} + (-1)^k
\sum_{i=1}^{n-2} \xi_i^{2(n-2-k)-1}
(1 + \lambda \xi_i^2)^{-1}
\prod_{\stackrel{\scriptstyle j=1}{(j \neq i)}}^{n-2} 
( \xi_i^2 - \xi_j^2)^{-1} ( d \tilde{\psi}_i - 2 A_i ).
\eeq
For $\mu \neq n-1$,
\beq
P_{\mu}= \frac{X_{\mu}(x_{\mu})}
{\displaystyle (x_{\mu})^{2m_{n-1}+2}
\prod_{i=1}^{n-2}( x_{\mu}^2 - \xi_i^2) \tilde{U}_{\mu}},
\qquad
\tilde{U}_{\mu} = 
\prod_{\stackrel{\scriptstyle \nu=1}{(\nu \neq \mu, n-1)}}^n
(x_{\mu}^2 - x_{\nu}^2),
\eeq
\beq
X_{\mu}(x_{\mu})
= (-1)^{(1/2)D} 2M i x_{\mu} \delta_{\mu, n}
- (1 + \lambda x_{\mu}^2) (x_{\mu})^{2m_{n-1}+2}
\prod_{i=1}^{n-2} ( x_{\mu}^2 - \xi_i^2)^{m_i+1}.
\eeq
Note that
\beq
g^{(i)} = (-1)^n (1 + \lambda \xi_i^2)^{-1} \xi_i^{-2}
\prod_{\stackrel{\scriptstyle j=1}{(j \neq i)}}^{n-1}
(\xi_i^2 - \xi_j^2)^{-1} d \Sigma_{i,(m_i)}^2.
\eeq
For $i \neq n-1$,
\beq
\tilde{g}^{(i)}
:= - \xi_i^2 \lim_{\xi_{n-1} \rightarrow 0} g^{(i)} 
= (-1)^{n-1} ( 1 + \lambda \xi_i^2)^{-1} \xi_i^{-2}
\prod_{\stackrel{\scriptstyle j=1}{(j \neq i)}}^{n-2}
( \xi_i^2 - \xi_j^2)^{-1} d \Sigma_{i,(m_i)}^2.
\eeq

In the $\xi_{n-1} \rightarrow 0$ limit,
the metric \eqref{evng} becomes
\beq \label{ggen0}
\begin{split}
g &= \sum_{\stackrel{\scriptstyle \mu=1}{(\mu \neq n-1)}}^n
\frac{d x_{\mu}^2}{P_{\mu}}
+ \sum_{\stackrel{\scriptstyle \mu=1}{(\mu \neq n-1)}}^n
P_{\mu}
\left[ \sum_{k=0}^{n-2} 
\tilde{\sigma}_k( \hat{x}_{\mu}) \tilde{\theta}_k \right]^2 \cr
&+ \sum_{i=1}^{n-2}  
\prod_{\stackrel{\scriptstyle \mu=1}{(\mu \neq n-1)}}^{n} 
(x_{\mu}^2 - \xi_i^2) \tilde{g}^{(i)}
- \hat{\sigma}_{n-1}
\left( \prod_{j=1}^{n-2} \xi_j^{-2} \right) 
d \tilde{\Omega}_{n-1,(2m_{n-1}+2)}^2,
\end{split}
\eeq
with
\beq
N = n-2, \qquad m^{(0)} = 2m_{n-1} + 2 ,\qquad
\varepsilon=0.
\eeq
The base space is given by
\beq
B = M^{(1)} \times M^{(2)} \times \dotsm
M^{(n-2)} \times M^{(0)}
= \mathbb{CP}^{m_1} \times
\mathbb{CP}^{m_2} \times
\dotsm \times \mathbb{CP}^{m_{n-2}}
\times S^{2m_{n-1}+2},
\eeq
and the fiber over $B$ is a $2(n-1)$-dimensional
Kerr-de Sitter space.

By replacing $n-1$ with $n$, we obtain \eqref{evng}
for even $m^{(0)}$. Hence 
we have shown that
the even dimensional
case of \eqref{evng} represents the even dimensional 
Kerr-de Sitter black hole 
with partially equal angular momenta
and with some zero angular momenta.

\section{Lichnerowicz operator}

\subsection{General type}

We evaluate the Lichnerowicz operator $\Delta_L h_{AB}$. 
The non-zero components are calculated
as follows:
\begin{flushleft}
$\bullet$~\textbf{vector} 
\end{flushleft}
\begin{equation} \label{gveccomp}
\Delta_L h_{n+\mu,(\hat{\alpha},i)}
= \frac{2 \xi_i \sqrt{P_{\mu}}}{x_{\mu}^2-\xi_i^2}
\left( \prod_{\nu=1}^n (x_\nu^2-\xi_i^2) \right)^{-1/2}
(\mathcal{D}^{(i)}_{\beta} h_{(m_i+\beta,i),(\hat{\alpha},i)}-
\mathcal{D}^{(i)}_{m_i+\beta} h_{(\beta,i),(\hat{\alpha},i)}),
\end{equation}
\begin{flushleft}
$\bullet$~\textbf{tensor} 
\end{flushleft}
\begin{eqnarray} \label{gtens1}
&\Delta_L& h_{(\alpha,i),(\beta,i)}
=\Box^{(F)} h_{(\alpha,i),(\beta,i)}
+2 \sum_{\mu} \frac{\xi_i \sqrt{P_{\mu}}}
{x_{\mu}^2-\xi_i^2} e_{n+\mu}
( h_{(m_i+\alpha,i),(\beta,i)}
+h_{(\alpha,i),(m_i+\beta,i)})\nonumber\\
&-&\sum_{(\hat{\gamma},j)(j\ne i)} 
\frac{1}{\prod_{\mu=1}^n (x_{\mu}^2-\xi_j^2)}
(\mathcal{D}^{(j)}_{\hat{\gamma}}
\bar{e}^{(j)}_{\hat{\gamma}}h_{(\alpha,i),(\beta,i)})
-2 \sum_{j, \mu} \frac{m_j x_{\mu} \sqrt{P_{\mu}}}
{x_{\mu}^2-\xi_j^2}
(e_{\mu}h_{(\alpha,i),(\beta,i)})\nonumber\\
&-& \sum_a \frac{1}{\sigma_n}
( D_a^{(0)} \tilde{e}_a h_{(\alpha, i),(\beta,i)})
- \sum_{\mu} \frac{m^{(0)} \sqrt{P_{\mu}}}{x_{\mu}}
( e_{\mu} h_{(\alpha,i), (\beta,i)}) \nonumber\\
&-&\frac{1}{\prod_{\mu=1}^n (x_{\mu}^2-\xi_i^2)}
\left( \sum_{\hat{\gamma}}
\mathcal{D}^{(i)}_{\hat{\gamma}}
\mathcal{D}^{(i)}_{\hat{\gamma}}h_{(\alpha,i),(\beta,i)}
+2 \sum_{\hat{\gamma}, \hat{\delta}}
\tilde{R}^{(i)}_{\alpha \hat{\gamma} \beta \hat{\delta}}
h_{(\hat{\gamma},i), (\hat{\delta},i)}
\right) \nonumber\\
&+&4 \sum_{\mu} \frac{\xi_i^2 P_{\mu}}
{(x_{\mu}^2-\xi_i^2)^2}
h_{(\alpha,i),(\beta,i)}
+4 \sum_{\mu,j} \frac{\xi_i \xi_j P_{\mu}}
{(x_{\mu}^2-\xi_i^2)(x_{\mu}^2-\xi_j^2)}
h_{(m_i+\alpha,i),(m_i+\beta,i)}\nonumber\\
&+&2 \Lambda h_{(\alpha,i),(\beta,i)},
\end{eqnarray}
\begin{eqnarray} \label{gtens2}
&\Delta_L& h_{(\alpha,i),(m_i+\beta,i)}
=\Box^{(F)} h_{(\alpha,i),(m_i+\beta,i)}
+2 \sum_{\mu} \frac{\xi_i \sqrt{P_{\mu}}}
{x_{\mu}^2-\xi_i^2} e_{n+\mu}
( h_{(m_i+\alpha,i),(m_i+\beta,i)}
-h_{(\alpha,i),(\beta,i)})\nonumber\\
&-&\sum_{(\hat{\gamma},j)(j\ne i)} 
\frac{1}{\prod_{\mu=1}^n (x_{\mu}^2-\xi_j^2)}
(\mathcal{D}^{(j)}_{\hat{\gamma}}
\bar{e}^{(j)}_{\hat{\gamma}}
h_{(\alpha,i),(m_i+\beta,i)})
-2 \sum_{j, \mu} \frac{m_j x_{\mu} \sqrt{P_{\mu}}}
{x_{\mu}^2-\xi_j^2}
(e_{\mu}h_{(\alpha,i),(m_i+\beta,i)})\nonumber\\
&-& \sum_a \frac{1}{\sigma_n} 
( D_a^{(0)} \tilde{e}_a h_{(\alpha, i), (m_i+\beta,i)} )
- \sum_{\mu} \frac{m^{(0)} \sqrt{P_{\mu}}}{x_{\mu}}
( e_{\mu} h_{(\alpha,i),(m_i+\beta,i)}) \nonumber\\
&-&\frac{1}{\prod_{\mu=1}^n (x_{\mu}^2-\xi_i^2)}
\left( \sum_{\hat{\gamma}}
\mathcal{D}^{(i)}_{\hat{\gamma}}
\mathcal{D}^{(i)}_{\hat{\gamma}}h_{(\alpha,i),(m_i+\beta,i)}
+2 \sum_{\hat{\gamma}, \hat{\delta}}
\tilde{R}^{(i)}_{\alpha, \hat{\gamma},m_i
+ \beta, \hat{\delta}}h_{(\hat{\gamma},i), (\hat{\delta},i)}
\right) \nonumber\\
&+&4 \sum_{\mu} \frac{\xi_i^2 P_{\mu}}{(x_{\mu}^2-\xi_i^2)^2}
h_{(\alpha,i),(m_i+\beta,i)}
-4 \sum_{\mu,j} \frac{\xi_i \xi_j P_{\mu}}
{(x_{\mu}^2-\xi_i^2)(x_{\mu}^2-\xi_j^2)}
h_{(m_i+\alpha,i),(\beta,i)}\nonumber\\
&+&2 \Lambda h_{(\alpha,i),(m_i+\beta,i)},
\end{eqnarray}
\begin{eqnarray} \label{gtens3}
&\Delta_L& h_{(m_i+\alpha,i),(m_i+\beta,i)}
=\Box^{(F)} h_{(m_i+\alpha,i),(m_i+\beta,i)}
-2 \sum_{\mu} \frac{\xi_i \sqrt{P_{\mu}}}
{x_{\mu}^2-\xi_i^2} e_{n+\mu}
( h_{(\alpha,i),(m_i+\beta,i)}
+h_{(m_i+\alpha,i),(\beta,i)})\nonumber\\
&-&\sum_{(\hat{\gamma},j)(j\ne i)} 
\frac{1}{\prod_{\mu=1}^n (x_{\mu}^2-\xi_j^2)}
(\mathcal{D}^{(j)}_{\hat{\gamma}}
\bar{e}^{(j)}_{\hat{\gamma}}h_{(m_i+\alpha,i),(m_i+\beta,i)})
-2 \sum_{j, \mu} \frac{m_j x_{\mu} \sqrt{P_{\mu}}}
{x_{\mu}^2-\xi_j^2}
(e_{\mu}h_{(m_i+\alpha,i),(m_i+\beta,i)})\nonumber\\
&-& \sum_a \frac{1}{\sigma_n}
( D_a^{(0)} \tilde{e}_a h_{(m_i+\alpha,i),(m_i+\beta,i)})
- \sum_{\mu} \frac{m^{(0)} \sqrt{P_{\mu}}}{x_{\mu}}
( e_{\mu} h_{(m_i+\alpha,i), ( m_i+\beta,i)} ) \nonumber\\
&-&\frac{1}{\prod_{\mu=1}^n (x_{\mu}^2-\xi_i^2)}
\left( \sum_{\hat{\gamma}}
\mathcal{D}^{(i)}_{\hat{\gamma}}
\mathcal{D}^{(i)}_{\hat{\gamma}}h_{(m_i+\alpha,i),(m_i+\beta,i)}
+2 \sum_{\hat{\gamma}, \hat{\delta}}
\tilde{R}^{(i)}_{m_i+\alpha, \hat{\gamma},m_i
+ \beta, \hat{\delta}}h_{(\hat{\gamma},i), (\hat{\delta},i)}
\right) \nonumber\\
&+&4 \sum_{\mu} \frac{\xi_i^2 P_{\mu}}{(x_{\mu}^2-\xi_i^2)^2}
h_{(m_i+\alpha,i),(m_i+\beta,i)}
+4 \sum_{\mu,j} \frac{\xi_i \xi_j P_{\mu}}
{(x_{\mu}^2-\xi_i^2)(x_{\mu}^2-\xi_j^2)}
h_{(\alpha,i),(\beta,i)}\nonumber\\
&+&2 \Lambda h_{(m_i+\alpha,i),(m_i+\beta,i)},
\end{eqnarray}

\begin{eqnarray}
&\Delta_L& h_{ab}=\Box^{(F)} h_{ab}
-\sum_{(\hat{\alpha},i)}
\frac{1}{\prod_{\mu=1}^n (x_{\mu}^2-\xi_i^2)}
(\mathcal{D}^{(i)}_{\hat{\alpha}} 
\bar{e}^{(i)}_{\hat{\alpha}}h_{ab})\nonumber\\
&-& 2 \sum_{i, \mu} \frac{m_i x_{\mu} \sqrt{P_{\mu}}}
{x_{\mu}^2-\xi_i^2}(e_{\mu}h_{ab})
-\frac{1}{\sigma_n} \left( 
\sum_{d} D^{(0)}_{d} D^{(0)}_{d}h_{ab}+2 \sum_{d,e}
\tilde{R}_{adbe}h_{de} \right)\nonumber\\
&-&m^{(0)} \sum_{\mu} \frac{\sqrt{P_{\mu}}}{x_{\mu}}
(e_{\mu}h_{ab})+2 \Lambda h_{ab},
\end{eqnarray}
where $\Box^{(F)}$ represents the scalar Laplacian 
on the fiber space. Explicitly, putting
$ ( e_{\hat{\mu}} )=(e_{\mu}, e_{n+\mu})$ we have\footnote{
In the special case, we put $(e_{\hat{\mu}})
=(\hat{e}_{\mu}, \hat{e}_{n+\mu},
\hat{e}_{2n+1} )$.}
\begin{equation}
\Box^{(F)} f=-\sum_{\hat{\mu}} e_{\hat{\mu}}(e_{\hat{\mu}}f)
+\sum_{\hat{\mu},\hat{\nu}}(e_{\hat{\nu}}f)
\omega_{\hat{\nu} \hat{\mu}}(e_{\hat{\mu}}).
\end{equation}

\subsection{Special type}

The non-zero components of
$\Delta_L h_{AB}$ are calculated as follows:
\begin{flushleft}
$\bullet$~\textbf{vector} 
\end{flushleft}
\begin{equation}
\Delta_L h_{2n+1,(\hat{\alpha},i)}
=- \frac{2 \sqrt{S}}{\xi_i}
\left( \prod_{\nu=1}^n (x_\nu^2-\xi_i^2) \right)^{-1/2}
(\mathcal{D}^{(i)}_{\beta} h_{(m_i+\beta,i),(\hat{\alpha},i)}-
\mathcal{D}^{(i)}_{m_i+\beta} h_{(\beta,i),(\hat{\alpha},i)})
\end{equation}
together with \eqref{gveccomp}.
\begin{flushleft}
$\bullet$~\textbf{tensor} 
\end{flushleft}
In the formulas \eqref{gtens1}, \eqref{gtens2} and \eqref{gtens3}, 
the terms in the third line
\begin{equation}
- \sum_{a} \frac{1}{\sigma_n} 
( D_a^{(0)} \tilde{e}_a h_{(\cdot, i), ( \cdot, i)})
- \sum_{\mu} \frac{m^{(0)} \sqrt{P_{\mu}}}{x_{\mu}}
( e_{\mu} h_{(\cdot,i),(\cdot,i)} )
\end{equation}
are dropped, and
the vector fields  
$\{ e_{\mu},e_{n+\mu} \}$ are replaced by 
$\{ \hat{e}_{\mu},\hat{e}_{n+\mu} \}$. 
The following new terms are added to the equations:
\begin{eqnarray}
&-&\frac{2 \sqrt{S}}{\xi_i} \hat{e}_{2n+1}
(h_{(m_i+\alpha,i),(\beta,i)}+h_{(\alpha,i),(m_i+\beta,i)})+
\frac{4 S}{\xi_i^2} h_{(\alpha,i),(\beta,i)}\nonumber\\
&+&4\sum_{j} \frac{S}{\xi_i \xi_j}
h_{(m_i+\alpha,i),(m_i+\beta,i)}~~~~~ \mbox{to \eqref{gtens1}}~,
\end{eqnarray}
\begin{eqnarray}
&-&\frac{2 \sqrt{S}}{\xi_i} 
\hat{e}_{2n+1}(h_{(m_i+\alpha,i),(m_i+\beta,i)}
-h_{(\alpha,i),(\beta,i)})+
\frac{4 S}{\xi_i^2} h_{(\alpha,i),(m_i+\beta,i)}\nonumber\\
&-&4\sum_{j} \frac{S}{\xi_i \xi_j}
h_{(m_i+\alpha,i),(\beta,i)}~~~~~\mbox{to \eqref{gtens2}}~,
\end{eqnarray}
\begin{eqnarray}
& &\frac{2 \sqrt{S}}{\xi_i} 
\hat{e}_{2n+1}(h_{(\alpha,i),(m_i+\beta,i)}
+h_{(m_i+\alpha,i),(\beta,i)})
+\frac{4 S}{\xi_i^2} 
h_{(m_i+\alpha,i),(m_i+\beta,i)}\nonumber\\
&+&4\sum_{j} \frac{S}{\xi_i \xi_j}
h_{(\alpha,i),(\beta,i)}~~~~~\mbox{ to \eqref{gtens3}}~.
\end{eqnarray}

\end{document}